\documentclass[english]{article}
\usepackage{lmodern}
\usepackage[T1]{fontenc}
\usepackage[latin9]{inputenc}
\usepackage{babel}
\usepackage{amsmath}
\usepackage{amssymb}
\usepackage{graphicx}
\usepackage{xargs}[2008/03/08]
\usepackage[unicode=true,pdfusetitle,
 bookmarks=true,bookmarksnumbered=false,bookmarksopen=false,
 breaklinks=false,pdfborder={0 0 1},backref=false,colorlinks=false]
 {hyperref}

\makeatletter

\providecommand{\tabularnewline}{\\}

\usepackage[]{aaai23}  % DO NOT CHANGE THIS
\usepackage{times}  % DO NOT CHANGE THIS
\usepackage{helvet}  % DO NOT CHANGE THIS
\usepackage{courier}  % DO NOT CHANGE THIS
\usepackage{natbib}  % DO NOT CHANGE THIS AND DO NOT ADD ANY OPTIONS TO IT
\usepackage{caption} % DO NOT CHANGE THIS AND DO NOT ADD ANY OPTIONS TO IT
\usepackage{algorithm}
\usepackage{algorithmic}

\usepackage{newfloat}
\usepackage{listings}
\DeclareCaptionStyle{ruled}{labelfont=normalfont,labelsep=colon,strut=off} % DO NOT CHANGE THIS
\floatstyle{ruled}
\newfloat{listing}{tb}{lst}{}
\floatname{listing}{Listing}
\title{Cross Project Software Vulnerability Detection via Domain Adaptation
and Max-Margin Principle}

\author{
Van Nguyen\thanks{\emph{The corresponding author. The initial work was done at Monash University, Australia before Van Nguyen joined Adelaide University, Australia. This research was supported under the Defence Science and Technology Group\textquoteright s Next Generation Technologies Program, Australia.}}$^1$~~~
Trung Le$^2$~~~
Chakkrit Tantithamthavorn$^2$~~~ \\
\textbf{
John Grundy$^2$~~~
Hung Nguyen$^1$~~~
Dinh Phung$^2$~~~}
\smallskip 
{\normalfont
\\
$^1$Adelaide University, Australia
\\
$^2$Monash University, Australia
\\
\texttt{\{khacvan.nguyen,hung.nguyen\}@adelaide.edu.au},
\\
\texttt{\{trunglm,chakkrit,john.grundy,dinh.phung\}@monash.edu}
}
}

\begin{document}

\newcommand{\sidenote}[1]{\marginpar{\small \emph{\color{Medium}#1}}}

\global\long\def\se{\hat{\text{se}}}%

\global\long\def\interior{\text{int}}%

\global\long\def\boundary{\text{bd}}%

\global\long\def\ML{\textsf{ML}}%

\global\long\def\GML{\mathsf{GML}}%

\global\long\def\HMM{\mathsf{HMM}}%

\global\long\def\support{\text{supp}}%

\global\long\def\new{\text{*}}%

\global\long\def\stir{\text{Stirl}}%

\global\long\def\head{\text{head}}%

\global\long\def\Concat{\text{Concat}}%

\global\long\def\LayerNormalization{\text{LayerNormalization}}%

\global\long\def\Attention{\text{Attention}}%

\global\long\def\MultiHeadAttention{\text{MultiHead-Attention}}%

\global\long\def\mA{\mathcal{A}}%

\global\long\def\mB{\mathcal{B}}%

\global\long\def\mF{\mathcal{F}}%

\global\long\def\mK{\mathcal{K}}%

\global\long\def\mH{\mathcal{H}}%

\global\long\def\mX{\mathcal{X}}%

\global\long\def\mZ{\mathcal{Z}}%

\global\long\def\mS{\mathcal{S}}%

\global\long\def\Ical{\mathcal{I}}%

\global\long\def\mT{\mathcal{T}}%

\global\long\def\Pcal{\mathcal{P}}%

\global\long\def\dist{d}%

\global\long\def\HX{\entro\left(X\right)}%
 
\global\long\def\entropyX{\HX}%

\global\long\def\HY{\entro\left(Y\right)}%
 
\global\long\def\entropyY{\HY}%

\global\long\def\HXY{\entro\left(X,Y\right)}%
 
\global\long\def\entropyXY{\HXY}%

\global\long\def\mutualXY{\mutual\left(X;Y\right)}%
 
\global\long\def\mutinfoXY{\mutualXY}%

\global\long\def\given{\mid}%

\global\long\def\gv{\given}%

\global\long\def\goto{\rightarrow}%

\global\long\def\asgoto{\stackrel{a.s.}{\longrightarrow}}%

\global\long\def\pgoto{\stackrel{p}{\longrightarrow}}%

\global\long\def\dgoto{\stackrel{d}{\longrightarrow}}%

\global\long\def\lik{\mathcal{L}}%

\global\long\def\logll{\mathit{l}}%

\global\long\def\vectorize#1{\boldsymbol{#1}}%

\global\long\def\vt#1{\mathbf{#1}}%

\global\long\def\gvt#1{\boldsymbol{#1}}%

\global\long\def\idp{\ \bot\negthickspace\negthickspace\bot\ }%
 
\global\long\def\cdp{\idp}%

\global\long\def\das{\triangleq}%

\global\long\def\id{\mathbb{I}}%

\global\long\def\idarg#1#2{\id\left\{  #1,#2\right\}  }%

\global\long\def\iid{\stackrel{\text{iid}}{\sim}}%

\global\long\def\bzero{\vt 0}%

\global\long\def\bone{\mathbf{1}}%

\global\long\def\boldm{\boldsymbol{m}}%

\global\long\def\be{\boldsymbol{e}}%

\global\long\def\bff{\vt f}%

\global\long\def\ba{\boldsymbol{a}}%

\global\long\def\bb{\boldsymbol{b}}%

\global\long\def\bc{\boldsymbol{c}}%

\global\long\def\bB{\boldsymbol{B}}%

\global\long\def\bx{\boldsymbol{x}}%

\global\long\def\bl{\boldsymbol{l}}%

\global\long\def\bu{\boldsymbol{u}}%

\global\long\def\bo{\boldsymbol{o}}%

\global\long\def\bh{\boldsymbol{h}}%

\global\long\def\bs{\boldsymbol{s}}%

\global\long\def\bz{\boldsymbol{z}}%

\global\long\def\xnew{y}%

\global\long\def\bxnew{\boldsymbol{y}}%

\global\long\def\bX{\boldsymbol{X}}%

\global\long\def\tbx{\tilde{\bx}}%

\global\long\def\by{\boldsymbol{y}}%

\global\long\def\bY{\boldsymbol{Y}}%

\global\long\def\bZ{\boldsymbol{Z}}%

\global\long\def\bU{\boldsymbol{U}}%

\global\long\def\bv{\boldsymbol{v}}%

\global\long\def\bn{\boldsymbol{n}}%

\global\long\def\bV{\boldsymbol{V}}%

\global\long\def\bI{\boldsymbol{I}}%

\global\long\def\bw{\vt w}%

\global\long\def\balpha{\gvt{\alpha}}%

\global\long\def\bbeta{\gvt{\beta}}%

\global\long\def\bmu{\gvt{\mu}}%

\global\long\def\btheta{\boldsymbol{\theta}}%

\global\long\def\bsigma{\boldsymbol{\sigma}}%

\global\long\def\blambda{\boldsymbol{\lambda}}%

\global\long\def\bgamma{\boldsymbol{\gamma}}%

\global\long\def\bpsi{\boldsymbol{\psi}}%

\global\long\def\bphi{\boldsymbol{\phi}}%

\global\long\def\bPhi{\boldsymbol{\Phi}}%

\global\long\def\bpi{\boldsymbol{\pi}}%

\global\long\def\bomega{\boldsymbol{\omega}}%

\global\long\def\bepsilon{\boldsymbol{\epsilon}}%

\global\long\def\btau{\boldsymbol{\tau}}%

\global\long\def\realset{\mathbb{R}}%

\global\long\def\realn{\realset^{n}}%

\global\long\def\integerset{\mathbb{Z}}%

\global\long\def\natset{\integerset}%

\global\long\def\integer{\integerset}%

\global\long\def\natn{\natset^{n}}%

\global\long\def\rational{\mathbb{Q}}%

\global\long\def\rationaln{\rational^{n}}%

\global\long\def\complexset{\mathbb{C}}%

\global\long\def\comp{\complexset}%

\global\long\def\compl#1{#1^{\text{c}}}%

\global\long\def\and{\cap}%

\global\long\def\compn{\comp^{n}}%

\global\long\def\comb#1#2{\left({#1\atop #2}\right) }%

\global\long\def\nchoosek#1#2{\left({#1\atop #2}\right)}%

\global\long\def\param{\vt w}%

\global\long\def\Param{\Theta}%

\global\long\def\meanparam{\gvt{\mu}}%

\global\long\def\Meanparam{\mathcal{M}}%

\global\long\def\meanmap{\mathbf{m}}%

\global\long\def\logpart{A}%

\global\long\def\simplex{\Delta}%

\global\long\def\simplexn{\simplex^{n}}%

\global\long\def\dirproc{\text{DP}}%

\global\long\def\ggproc{\text{GG}}%

\global\long\def\DP{\text{DP}}%

\global\long\def\P{\text{P}}%

\global\long\def\R{\text{R}}%

\global\long\def\FNR{\text{FNR}}%

\global\long\def\FPR{\text{FPR}}%

\global\long\def\TP{\text{TP}}%

\global\long\def\FP{\text{FP}}%

\global\long\def\FN{\text{FN}}%

\global\long\def\TN{\text{TN}}%

\global\long\def\F{\text{F1}}%

\global\long\def\ndp{\text{nDP}}%

\global\long\def\hdp{\text{HDP}}%

\global\long\def\gempdf{\text{GEM}}%

\global\long\def\Gumbel{\text{Gumbel}}%

\global\long\def\Uniform{\text{Uniform}}%

\global\long\def\Mult{\text{Mult}}%

\global\long\def\rfs{\text{RFS}}%

\global\long\def\bernrfs{\text{BernoulliRFS}}%

\global\long\def\poissrfs{\text{PoissonRFS}}%

\global\long\def\grad{\gradient}%
 
\global\long\def\gradient{\nabla}%

\global\long\def\partdev#1#2{\partialdev{#1}{#2}}%
 
\global\long\def\partialdev#1#2{\frac{\partial#1}{\partial#2}}%

\global\long\def\partddev#1#2{\partialdevdev{#1}{#2}}%
 
\global\long\def\partialdevdev#1#2{\frac{\partial^{2}#1}{\partial#2\partial#2^{\top}}}%

\global\long\def\closure{\text{cl}}%

\global\long\def\cpr#1#2{\Pr\left(#1\ |\ #2\right)}%

\global\long\def\var{\text{Var}}%

\global\long\def\Var#1{\text{Var}\left[#1\right]}%

\global\long\def\cov{\text{Cov}}%

\global\long\def\Cov#1{\cov\left[ #1 \right]}%

\global\long\def\COV#1#2{\underset{#2}{\cov}\left[ #1 \right]}%

\global\long\def\corr{\text{Corr}}%

\global\long\def\sst{\text{T}}%

\global\long\def\SST{\sst}%

\global\long\def\ess{\mathbb{E}}%

\global\long\def\Ess#1{\ess\left[#1\right]}%

\newcommandx\ESS[2][usedefault, addprefix=\global, 1=]{\underset{#2}{\ess}\left[#1\right]}%

\global\long\def\fisher{\mathcal{F}}%

\global\long\def\bfield{\mathcal{B}}%
 
\global\long\def\borel{\mathcal{B}}%

\global\long\def\bernpdf{\text{Bernoulli}}%

\global\long\def\betapdf{\text{Beta}}%

\global\long\def\dirpdf{\text{Dir}}%

\global\long\def\gammapdf{\text{Gamma}}%

\global\long\def\gaussden#1#2{\text{Normal}\left(#1, #2 \right) }%

\global\long\def\gauss{\mathbf{N}}%

\global\long\def\gausspdf#1#2#3{\text{Normal}\left( #1 \lcabra{#2, #3}\right) }%

\global\long\def\multpdf{\text{Mult}}%

\global\long\def\poiss{\text{Pois}}%

\global\long\def\poissonpdf{\text{Poisson}}%

\global\long\def\pgpdf{\text{PG}}%

\global\long\def\wshpdf{\text{Wish}}%

\global\long\def\iwshpdf{\text{InvWish}}%

\global\long\def\nwpdf{\text{NW}}%

\global\long\def\niwpdf{\text{NIW}}%

\global\long\def\studentpdf{\text{Student}}%

\global\long\def\unipdf{\text{Uni}}%

\global\long\def\transp#1{\transpose{#1}}%
 
\global\long\def\transpose#1{#1^{\mathsf{T}}}%

\global\long\def\mgt{\succ}%

\global\long\def\mge{\succeq}%

\global\long\def\idenmat{\mathbf{I}}%

\global\long\def\trace{\mathrm{tr}}%

\global\long\def\argmax#1{\underset{_{#1}}{\text{argmax}} }%

\global\long\def\argmin#1{\underset{_{#1}}{\text{argmin}\ } }%

\global\long\def\diag{\text{diag}}%

\global\long\def\concat{\text{concat}}%

\global\long\def\softmax{\text{softmax}}%

\global\long\def\norm{}%

\global\long\def\spn{\text{span}}%

\global\long\def\vtspace{\mathcal{V}}%

\global\long\def\field{\mathcal{F}}%
 
\global\long\def\ffield{\mathcal{F}}%

\global\long\def\inner#1#2{\left\langle #1,#2\right\rangle }%
 
\global\long\def\iprod#1#2{\inner{#1}{#2}}%

\global\long\def\dprod#1#2{#1 \cdot#2}%

\global\long\def\norm#1{\left\Vert #1\right\Vert }%

\global\long\def\entro{\mathbb{H}}%

\global\long\def\entropy{\mathbb{H}}%

\global\long\def\Entro#1{\entro\left[#1\right]}%

\global\long\def\Entropy#1{\Entro{#1}}%

\global\long\def\mutinfo{\mathbb{I}}%

\global\long\def\relH{\mathit{D}}%

\global\long\def\reldiv#1#2{\relH\left(#1||#2\right)}%

\global\long\def\KL{KL}%

\global\long\def\KLdiv#1#2{\KL\left(#1\parallel#2\right)}%
 
\global\long\def\KLdivergence#1#2{\KL\left(#1\ \parallel\ #2\right)}%

\global\long\def\crossH{\mathcal{C}}%
 
\global\long\def\crossentropy{\mathcal{C}}%

\global\long\def\crossHxy#1#2{\crossentropy\left(#1\parallel#2\right)}%

\global\long\def\breg{\text{BD}}%

\global\long\def\lcabra#1{\left|#1\right.}%

\global\long\def\lbra#1{\lcabra{#1}}%

\global\long\def\rcabra#1{\left.#1\right|}%

\global\long\def\rbra#1{\rcabra{#1}}%

\maketitle

\begin{abstract}
Software vulnerabilities (SVs) have become a common, serious and crucial
concern due to the ubiquity of computer software. Many machine learning-based
approaches have been proposed to solve the software vulnerability
detection (SVD) problem. However, there are still  two open and significant
issues  for SVD in terms of i) learning automatic representations
to improve the predictive performance of SVD, and ii) tackling the
scarcity of labeled vulnerabilities datasets that conventionally need
laborious labeling effort by experts. In this paper, we propose a
novel end-to-end approach to tackle these two crucial issues. We first
exploit the automatic representation learning with deep domain adaptation
for software vulnerability detection. We then propose a novel cross-domain
kernel classifier leveraging the max-margin principle to significantly
improve the transfer learning process of software vulnerabilities
from labeled projects into unlabeled ones. The experimental results
on real-world software datasets show the superiority of our proposed
method over state-of-the-art baselines. In short, our method obtains a higher performance on F1-measure, the most important measure in SVD, from 1.83\% to 6.25\% compared to the second highest method in the used datasets. Our released source code samples are publicly available at \href{https://github.com/vannguyennd/dam2p}{https://github.com/vannguyennd/dam2p.}
\end{abstract}

\section{Introduction\label{sec:intro}}
\vspace{1mm}
%In the field of software security, 
Software vulnerabilities (SVs),
defined as specific flaws or oversights in software programs allowing
attackers to exploit the code base and potentially undertake dangerous
activities (e.g., exposing or altering sensitive information, disrupting,
degrading or destroying a system, or taking control of a program or
computer system) \cite{Dowd2006}, are very common and represent major security risks due
to the ubiquity of computer software. Detecting  and eliminating software vulnerabilities are hard as software development technologies and methodologies vary significantly between projects and products.
%Given the vast variety of technologies
%and different software development methodologies used, a great deal
%of computer software may naturally contain software vulnerabilities,
%hence making the problem of software vulnerability detection become
%a critical concern in computer security and software engineering.
The severity of the threat imposed by software vulnerabilities
(SVs) has significantly increased over the years causing significant
damages to companies and individuals. The worsening software vulnerability situation has necessitated
the development of automated advanced approaches and tools that can
efficiently and effectively detect SVs with a minimal level of human
intervention. To respond to this demand, many vulnerability detection
systems and methods, ranging from open source to commercial tools,
and from manual to automatic methods \cite{Neuhaus:2007:PVS,shin2011evaluating,yamaguchi2011vulnerability,Grieco2016,Li2016:VAV,KimWLO17,VulDeePecker2018,Duan2019,Cheng2019,Zhuang2020}
have been proposed and implemented.

Most previous work in software vulnerability detection (SVD)\cite{Neuhaus:2007:PVS,shin2011evaluating,yamaguchi2011vulnerability,Li2016:VAV,Grieco2016,KimWLO17}
has based primarily on handcrafted features which are manually
chosen by knowledgeable domain experts with possibly outdated experience
and underlying biases. In many situations, these handcrafted features 
do not generalize well. For example, features that work well in a
certain software project may not perform well in other projects \cite{Zimmermann2009}.
To alleviate the dependency on handcrafted features, the use of automatic
features in SVD has been studied recently \cite{VulDeePecker2018,jun_2018,Dam2018,Li2018SySeVR,Duan2019,Cheng2019,Zhuang2020}.
These works show the advantages of employing automatic features
over handcrafted features for software vulnerability
detection.

Another major challenging issue in SVD research is the scarcity of
labeled software projects to train models. The process of labeling
vulnerable source code is tedious, time-consuming, error-prone, and
can be very challenging even for domain experts. This has resulted
in few labeled projects compared with a vast volume of unlabeled ones.
Some recent approaches \cite{vannguyen2019dan,van-nguyen-dual-dan-2020,Liu2020CD-VulD}
have been proposed to solve this challenging problem with the aim
to transfer the learning of vulnerabilities from labeled source domains
to unlabeled target domains. Particularly, the methods in \citet{vannguyen2019dan,van-nguyen-dual-dan-2020}
learn domain-invariant features from the source code data of the source and
target domains by using the adversarial learning framework such as
Generative Adversarial Network (GAN) \cite{goodfellow2014generative}
while the method in \citet{Liu2020CD-VulD} consists of many subsequent
stages: i) pre-training a deep feature model for learning representation
of token sequences (i.e., source code data), ii) learning cross-domain
representations using a transformation to project token sequence embeddings
from (i) to a latent space, and iii) training a classifier from the
representations of the source domain data obtained from (ii). \emph{However,
none of these methods exploit the imbalanced nature of source code
projects for which the vulnerable  data points are significantly minor comparing
to non-vulnerable ones}.

On the other side, kernel methods are well-known for their ability
to deal with imbalanced datasets \cite{scholkopf2001,tax_svdd,Tsang05corevector,Tsang2007,trung2010,Van2014}.
In a nutshell, a \emph{domain of majority}, which is often a simple
geometric shape in a feature space (e.g., half-hyperplane \cite{scholkopf2001,Van2014}
or hypersphere \cite{tax_svdd,Tsang05corevector,Tsang2007,trung2010}
is defined to characterize the majority class (i.e., non-vulnerable
data). The key idea is that a simple domain of majority in the feature
space, when being mapped back to the input space, forms a set of contours
which can distinguish majority data from minority data.

In this paper, by leveraging learning domain-invariant features and
kernel methods with the max-margin principle, we propose \textbf{\emph{D}}\emph{omain
}\textbf{\emph{A}}\emph{daptation with }\textbf{\emph{M}}\emph{ax-}\textbf{\emph{M}}\emph{argin
}\textbf{\emph{P}}\emph{rinciple} (DAM2P) to efficiently transfer
learning from imbalanced labeled source domains to imbalanced unlabeled
target domains. Particularly, inspired by the max-margin principle
proven efficiently and effectively for learning from imbalanced data,
when learning domain-invariant features, we propose to learn a max-margin
hyperplane on the feature space to separate vulnerable and non-vulnerable
data. More specifically, we combine labeled source domain data and unlabeled
target domain data, and then learn a hyperplane to separate \emph{source domain
non-vulnerable from vulnerable data} and \emph{target domain data from the
origin} such that the margin is maximized. In addition, the margin
is defined as the minimization of the source domain and target domain margins in
which the \emph{source domain margin} is regarded as the minimal distance
from vulnerable data points to hyperplane \cite{Van2014}, while the
\emph{target domain margin} is regarded as the distance from the origin to
the hyperplane \cite{scholkopf2001}.

\vspace{1mm}
Our contributions can be summarized
as follows:
\begin{itemize}
\item We leverage learning domain-invariant features and kernel methods
with the max-margin principle to propose a novel approach named DAM2P
that can bridge the gap between the source and target domains on
a joint space while being able to tackle efficiently and effectively
the imbalanced nature of the source and target domains to significantly
improve the transfer learning process of SVs from labeled projects
into unlabeled ones.
\item We conduct extensive experiments on real-world software datasets consisting
of FFmpeg, LibTIFF, LibPNG, VLC, and Pidgin software projects. It
is worth noting that to demonstrate and compare the capability of
our proposed method and baselines in the transfer learning for SVD,
the datasets (FFmpeg, VLC, and Pidgin) from the \emph{multimedia application
domain }are used as the source domains whilst the datasets (LibPNG
and LibTIFF) from the \emph{image application domain} were used as
the target domains. The experimental results show that our method
significantly outperforms the baselines by a wide margin, especially
for the F1-measure, the most important measure used in  SVD problems.
\end{itemize}

\section{Related Work}

Automatic features in software vulnerability detection (SVD) has been
widely studied \cite{VulDeePecker2018,jun_2018,Dam2018,Li2018SySeVR,Duan2019,Cheng2019,Zhuang2020}
due to its advantages of employing automatic features over handcrafted
features. In particular, \citet{Dam2018} employed
a deep neural network to transform sequences of code tokens to vectorial
features that are further fed to a separate classifier; whereas \citet{VulDeePecker2018}
combined the learning of the vector representation and the training
of the classifier in a deep network. Advanced deep net architectures
have been investigated for SVD problem. \citet{Rebecca2018}
combined both recurrent neural networks (RNNs) and convolutional neural
networks (CNNs) for feature extraction from the embedded source code
representations while \citet{Zhuang2020} proposed a new model for
smart contract vulnerability detection based on a graph neural network
(GNN) \cite{KipfW16}.

Deep domain adaptation-based methods have been recently studied for
cross-domain SVD. Notably, \citet{vannguyen2019dan} proposed a novel architecture
and employed the adversarial learning framework (e.g., GAN) to learn
domain-invariant features that can be transferred from labeled source
to unlabeled target code project. \citet{van-nguyen-dual-dan-2020}
enhanced \citet{vannguyen2019dan} by proposing an elegant workaround
to combat the mode collapsing problem possibly faced in that work
due to the use of GAN. Finally, \citet{Liu2020CD-VulD} proposed a
multi-stage approach  with three sequential stages:
i) pre-training a deep model for learning representation of token
sequences (i.e., source code data), ii) learning cross-domain representations
using a transformation to project token sequence embeddings from (i)
to a latent space, iii) training a classifier from the representations
of the source domain data obtained from (ii). The trained classifier is then applied to the target domain.

In computer vision, deep domain adaptation (DA) has been intensively
studied and shown appealing performance in various transfer learning
tasks and applications, notably DDAN \cite{Ganin2015}, MMD \cite{long2015},
D2GAN \cite{duald-tunguyen2017}, DIRT-T \cite{shu2018a}, HoMM \cite{Hommchen2020},
and LAMDA \cite{LAMDAle21a}. Most of the introduced methods were
claimed, applied and showed the results for vision data. There is
no evidence that they can straightforwardly be applied for sequential
data such as in cross-domain SVD. In our paper, inspired from \citet{vannguyen2019dan},
we borrowed the principles of some well-known and state-of-the-art
methods, e.g., DDAN, MMD, D2GAN, DIRT-T, HoMM and LAMDA, and refactored
them using the CDAN architecture introduced in \citet{vannguyen2019dan}
for cross-domain SVD to compare with our proposed approach.

\section{Domain Adaptation with Max-Margin Principle\label{sec:Our-Approach}}

\subsection{Problem Formulation \label{subsec:The-Problem-Statement}}

Given a labeled source domain dataset $S=\left\{ \left(\bx_{1}^{S},y_{1}\right),\dots,\left(\bx_{N_{S}}^{S},y_{N_{S}}\right)\right\} $
where $y_{i}\in\left\{ 0,1\right\} $ (i.e., $1$: vulnerable code
and $0$: non-vulnerable code), let $\bx_{i}^{S}=\left[\bx_{i1}^{S},\dots,\bx_{iL}^{S}\right]$
be a code function represented as a sequence of $L$ embedding vectors.
We note that each embedding vector corresponds to a statement in the
code function. Similarly, the unlabeled target domain dataset $T=\left\{ \bx_{1}^{T},\dots,\bx_{N_{T}}^{T}\right\} $
consists of many code functions where each code function $\bx_{i}^{T}=\left[\bx_{i1}^{T},\dots,\bx_{iL}^{T}\right]$
is a sequence of $L$ embedding vectors.

\vspace{1mm}
Figure \ref{fig:An-example-of-processed-data} shows an example of the overall procedure for the source code data processing and embedding (\emph{please refer to the appendix for details}).

\begin{figure}[H]
\vspace{-2mm}
\begin{centering}
\includegraphics[width=0.9\columnwidth]{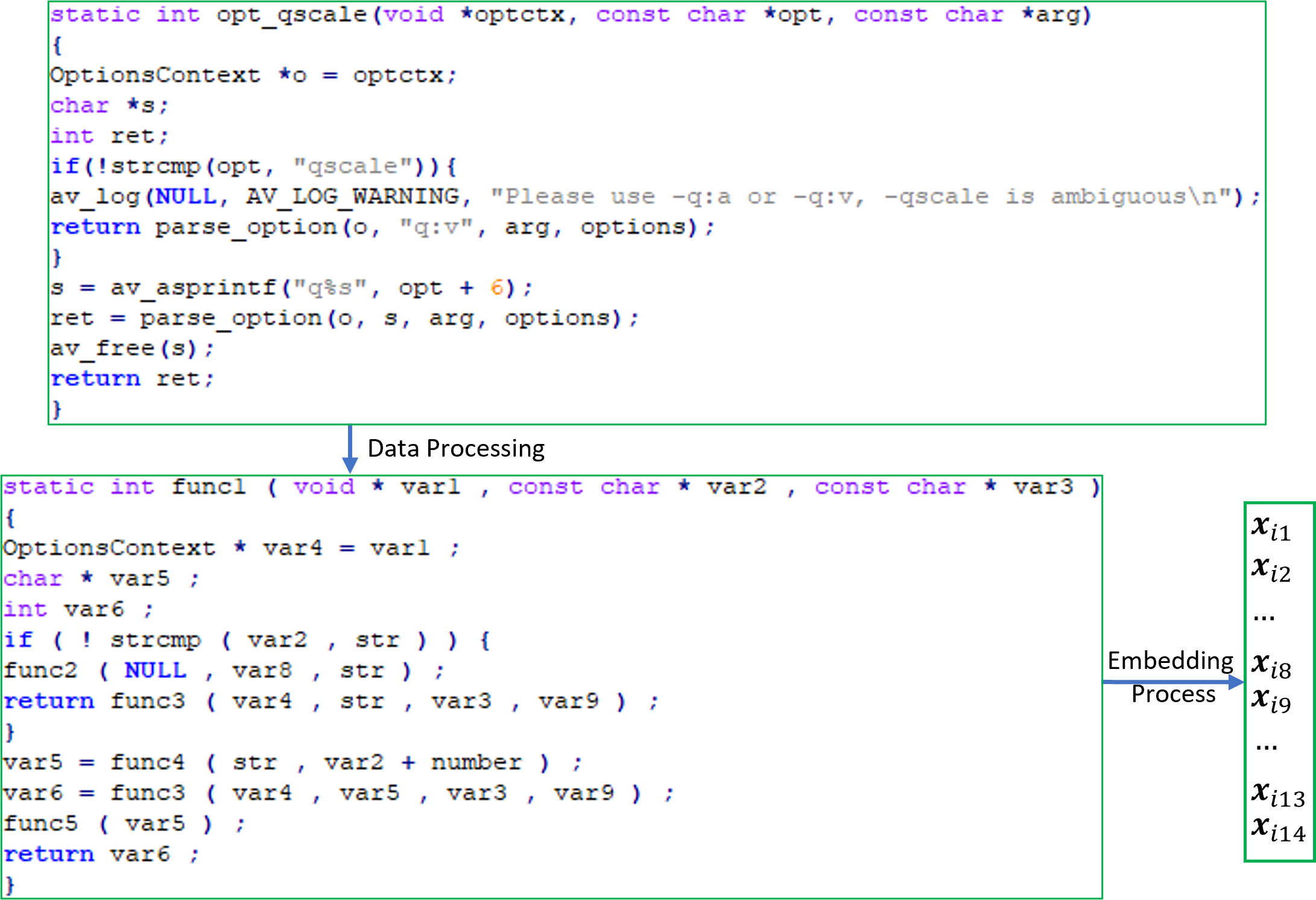}
\par\end{centering}\vspace{-2mm}
\caption{An example of the overall procedure for data processing and embedding of a source code function. After the data processing step (e.g., removing comments and non-ASCII characters), we use the embedding process to obtain the embedded
vectors for the source code statements.\label{fig:An-example-of-processed-data}}
\vspace{-3mm}
\end{figure}

In standard DA approaches, domain-invariant features
are learned on a joint space so that a classifier mainly trained based
on labeled source domain data can be transferred to predict well unlabeled
target domain data. The classifiers of interest are usually deep nets conducted
on top of domain-invariant features. In this work, by leveraging 
the kernel theory and the max-margin principle, we consider a kernel
machine on top of domain-invariant features, which is a hyperplane
on a feature space via a feature map $\phi$.

Inspired by the max-margin principle proven efficiency and effectiveness
for learning from imbalanced data, when learning domain-invariant
features, we propose to learn a max-margin hyperplane on the feature
space to separate vulnerable and non-vulnerable code data. More specifically,
we combine labeled source domain data and unlabeled target domain data, and then
learn a hyperplane to separate \emph{source domain non-vulnerable from vulnerable
data} and \emph{target domain data from the origin} such that the margin
is maximized. Here, the margin is defined as the minimization
of the source domain and target domain margins in which the \emph{source domain margin}
is defined as the minimum distance from vulnerable data points to
hyperplane \cite{Van2014} while the \emph{target domain margin} is defined
as the distance from the origin to the hyperplane \cite{scholkopf2001}.

\subsection{Our Proposed Approach DAM2P \label{subsec:Deep-Domain-Maximum-Margin}}

\subsubsection{Domain Adaptation for Domain-Invariant Features}

In what follows, we present the architecture of the generator $G$
and how to use an adversarial learning framework such as GAN \cite{goodfellow2014generative}
to learn domain-invariant features in a joint latent space specified
by $G$. To learn the automatic features for the sequential source
code data, inspired by \citet{VulDeePecker2018,vannguyen2019dan,van-nguyen-dual-dan-2020},
we apply a bidirectional recurrent neural network (bidirectional RNN)
to both the source and target domains. Given a source code function
$\bx$ in the source domain or the target domain, we denote the output
of the bidirectional RNN by $\mathcal{B}\left(\bx\right)$. We then
use some fully connected layers to connect the output layer of the
bidirectional RNN with the joint feature layer wherein we bridge the
gap between the source and target domains. The generator is consequently
the composition of the bidirectional RNN and the fully connected layers:
$G\left(\bx\right)=f\left(\mathcal{B}\left(\bx\right)\right)$ where
$f\left(\cdot\right)$ represents the map formed by the fully connected
layers.

Subsequently, to bridge the gap between the source and target domains
in the latent space, inspired by GAN \cite{goodfellow2014generative},
we use a domain discriminator $D$ to discriminate the source domain and
target domain data and train the generator $G$ to fool the discriminator
$D$ by making the source domain and target domain data indistinguishable in the
latent space. The relevant objective function is hence as follows:\vspace{-1mm}
\begin{align}
\mathcal{H}\left(G,D\right):= & \frac{1}{N_{S}}\sum_{i=1}^{N_{S}}\log D\left(G\left(\bx_{i}^{S}\right)\right)\nonumber \\
+ & \frac{1}{N_{T}}\sum_{i=1}^{N_{T}}\log\left[1-D\left(G\left(\bx_{i}^{T}\right)\right)\right]\label{eq:dan}
\end{align}

where we seek the optimal generator $G^{*}$ and the domain discriminator
$D^{*}$ by solving:
\[
G^{*}=\argmin G\mathcal{H}\left(G,D\right)\,\text{and}\,D^{*}=\argmax D\,\mathcal{H}\left(G,D\right)
\]

\subsubsection{Cross-domain kernel classifier}

To build up an efficient domain adaptation approach for source code
data which can tackle well the imbalanced nature of source code projects,
we leverage learning domain-invariant features with the max-margin
principle in the context of kernel machines to propose a novel cross-domain
kernel classifier named DAM2P. We construct a hyperplane on the feature
space: $\bw^{T}\phi(G(\bx))-\rho=0$ with the feature map $\phi$
and learn this hyperplane using the max-margin principle. More specifically,
we combine labeled source domain and unlabeled target domain data, and then learn
a hyperplane to separate \emph{source domain non-vulnerable from vulnerable
data} and \emph{target domain data from the origin} in such a way that the
margin is maximized.

\begin{figure*}[t]
\centering{}\includegraphics[width=1\textwidth]{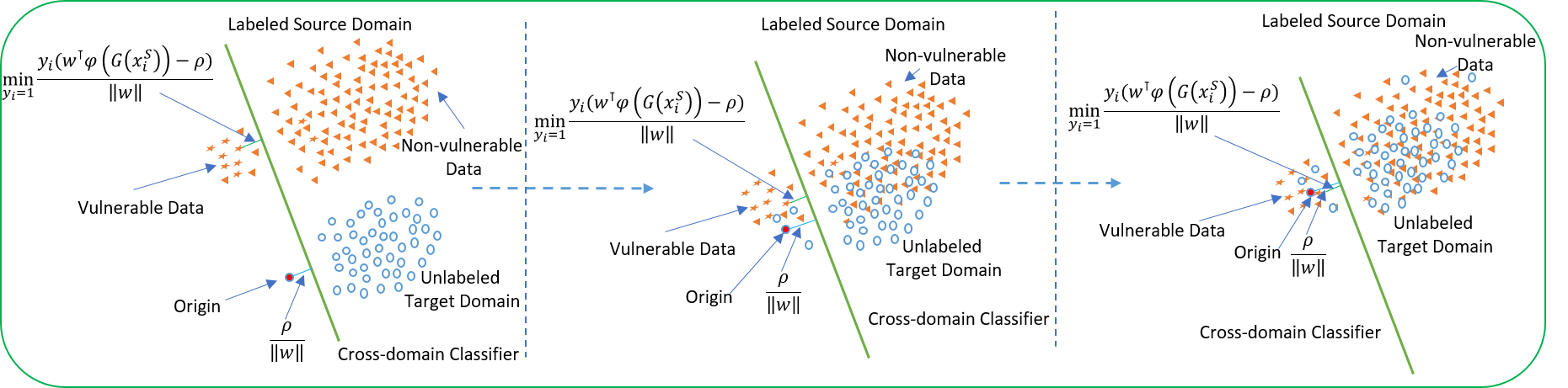}\vspace{-2mm}
\caption{The architecture of our cross-domain kernel classifier in
the feature space. By using our DAM2P method, we can gradually
bridge the gap between the source and target domains in the latent
space, while in the feature space our cross-domain kernel
classifier helps to distinguish the vulnerable and non-vulnerable
data. At the end, when the source and target domains are intermingled,
we can transfer our trained cross-domain kernel classifier to classify
the data of the target domain. \label{fig:The-architecture-of-cross-classifier}}
\vspace{-3mm}
\end{figure*}

It is worth noting that in this work, the margin is defined as the minimization of the \emph{source
}and \emph{target margins} in which the \emph{source margin} is the
minimal distance from the source domain vulnerable data points to the hyperplane \cite{Van2014},
while the \emph{target margin} is the distance from the origin to
the hyperplane \cite{scholkopf2001}. The overall architecture of
our proposed cross-domain kernel classifier in the feature space is
depicted in Figure \ref{fig:The-architecture-of-cross-classifier}.

Given the source domain dataset $S=\{(\bx_{1}^{S},y_{1}),\dots,(\bx_{N_{S}}^{S},y_{N_{S}})\}$
where $y_{i}=1,\,i=1,...,m$ and $y_{i}=0,\,i=m+1,...,N_{S}$ and
the target domain dataset $T=\{\bx_{1}^{T},\dots,\bx_{N_{T}}^{T}\}$,
we formulate the following optimization problem:

\begin{equation}
\underset{\bw,\rho}{\text{max}}\left(\text{min}\left\{ \underset{\text{source margin}}{\underbrace{\underset{y_{i}=1}{\text{min}}\left\{ \frac{y_{i}(\bw^{\top}\phi(G(\boldsymbol{x}_{i}^{S}))-\rho)}{\left\Vert \bw\right\Vert }\right\} }},\underset{\text{target margin}}{\underbrace{\frac{\rho}{\left\Vert \bw\right\Vert }}}\right\} \right)\label{eq:lm-ocsvm-original}
\end{equation}
subject to 
\begin{align*}
y_{i}(\bw^{\top}\phi(G(\boldsymbol{x}_{i}^{S}))-\rho) & \geq0,\,i=1,...,N_{S}\\
\bw^{\top}\phi(G(\boldsymbol{x}_{i}^{T}))\geq\rho, & \,i=1,...,N_{T}.
\end{align*}

In the optimization problem (\ref{eq:lm-ocsvm-original}), $\bw$ and
$\rho$ are the normal vector and the bias of the hyperplane and $\phi$
is a transformation from the joint latent space to the feature space,
while $G$ is the generator used to map the data of source and target
domains from the input space into the joint latent space. It occurs
that the margin is invariant if we scale $\bw,\rho$ by a factor $k>0$.
Hence without loosing of generality, we can assume that $\text{min}\left\{ \underset{y_{i}=1}{\textrm{min}}\left\{ y_{i}\bw^{\top}\phi(G(\boldsymbol{x}_{i}^{S})-\rho\right\} ,\rho\right\} =1$\footnote{This assumption is feasible because if $\left(\bw^{*},\rho^{*}\right)$
is the optimal solution, $\left(k\bw^{*},k\rho^{*}\right)$ with $k>0$
is also another optimal solution. Therefore, we can choose $k$ to
satisfy the assumption.}. The optimization problem (\ref{eq:lm-ocsvm-original}) can be rewritten
as follows:
\begin{equation}
\underset{w,\rho}{\textrm{min}}\frac{1}{2}\left\Vert \bw\right\Vert ^{2}\label{eq:ocsvm-lm-ocsvm-hard}
\end{equation}
subject to
\[
\begin{array}{l}
y_{i}(\bw^{\top}\phi(G(\boldsymbol{x}_{i}^{S}))-\rho)\geq0,\,i=1,...,m\\
y_{i}(\bw^{\top}\phi(G(\boldsymbol{x}_{i}^{S}))-\rho)\geq1,\,i=m+1,...,N_{S}\\
\bw^{\top}\phi(G(\boldsymbol{x}_{i}^{T}))\geq\rho,\,i=1,...,N_{T}
\end{array}
\]

We refer the above model as hard version of our proposed cross-domain
kernel classifier. To derive the soft version, we extend the optimization
problem in Eq. (\ref{eq:ocsvm-lm-ocsvm-hard}) by using the slack
variables as follows:
\begin{equation}
\underset{\bw,\rho}{\textrm{min}}\left(\frac{1}{2}\left\Vert \bw\right\Vert ^{2}+\frac{1}{N_{S}+N_{T}}\left(\sum_{i=1}^{N_{S}}\xi_{i}^{S}+\lambda\sum_{i=1}^{N_{T}}\xi_{i}^{T}\right)\right)\label{eq:ocsvm-l-ocsvm-soft}
\end{equation}
subject to
\[
\begin{array}{l}
y_{i}(\bw^{\top}\phi(G(\boldsymbol{x}_{i}^{S}))-\rho)\geq-\xi_{i}^{S},\,i=1,...,m\\
y_{i}(\bw^{\top}\phi(G(\boldsymbol{x}_{i}^{S}))-\rho)\geq1-\xi_{i}^{S},\,i=m+1,...,N_{S}\\
\bw^{\top}\phi(G(\boldsymbol{x}_{i}^{T}))\geq\rho-\xi_{i}^{T},\,i=1,...,N_{T}\\
\xi_{i}^{S}\geq0,\,i=1,...,N_{S};\xi_{i}^{T}\geq0,\,i=1,...,N_{T}.
\end{array}
\]
where $\lambda>0$ is the trade-off hyper-parameter representing the
weight of the information from the target domain contributing to the
cross-domain kernel classifier.

The primal form of the soft model optimization problem is hence of
the following form:
\begin{gather}
\underset{\bw,\rho}{\text{min}}\mathcal{L}(G,\bw,\rho)\label{eq:ocsvm-lm-ocsvm-soft-primal}
\end{gather}

where we have defined
\begin{align*}
\mathcal{L}(G,\bw,\rho):= & \frac{1}{2}\left\Vert \bw\right\Vert ^{2}+\frac{1}{N_{S}+N_{T}}\underset{i=1}{\overset{m}{\sum}}\max\left\{ 0,-z_{i}\right\} \\
+ & \frac{1}{N_{S}+N_{T}}\underset{i=m+1}{\overset{N_{S}}{\sum}}\max\left\{ 0,-z_{i}+1\right\} \\
+ & \frac{\lambda}{N_{S}+N_{T}}\underset{i=1}{\overset{N_{T}}{\sum}}\max\left\{ 0,-\bw^{\top}\phi(G(\boldsymbol{x}_{i}^{T}))+\rho\right\} 
\end{align*}
with $z_{i}=y_{i}\left(\bw^{\top}\phi(G(\boldsymbol{x}_{i}^{S})-\rho\right)$.
We use a random feature map \cite{Rahimi_NIPS2007} for the transformation
$\phi$ to map the representations (e.g., $G(\boldsymbol{x}_{i}^{S})$
and $G(\boldsymbol{x}_{i}^{T})$) from the latent space to a random
feature space. The formulation of $\phi$ on a specific $G(\boldsymbol{x}_{i})\in\mathbb{R}^{d}$
is as follows:
\[
\phi(G(\boldsymbol{x}_{i}))=[\frac{1}{\sqrt{K}}cos(\omega_{k}^{\top}G(\boldsymbol{x}_{i}),\frac{1}{\sqrt{K}}sin(\omega_{k}^{\top}G(\boldsymbol{x}_{i})]_{k=1}^{K}
\]

where $K$ consists of independent and identically distributed samples
$\omega_{1},...,\omega_{K}\in\mathbb{R}^{d}$ which are the Fourier
random elements.

We note that the use of a random feature map $\phi$ \cite{Rahimi_NIPS2007}
in conjunction with the cost-sensitive kernel machine of our proposed
cross-domain kernel classifier as mentioned in Eq. (\ref{eq:ocsvm-lm-ocsvm-soft-primal})
and a bidirectional recurrent neural network for the generator $G$
allows us to conveniently do back-propagation when training our proposed
approach. Combining the optimization problems in Eqs. (\ref{eq:dan}
and \ref{eq:ocsvm-lm-ocsvm-soft-primal}), we arrive at the final
objective function:
\begin{equation}
\mathcal{I}\left(G,D,\bw,\rho\right):=\mathcal{L}(G,\bw,\rho)+\alpha\mathcal{H}\left(G,D\right)\label{eq:final-objective-function}
\end{equation}
where $\alpha>0$ is the trade-off hyper-parameter. We seek the optimal
generator $G^{*}$, domain discriminator $D^{*}$, the normal vector
$\bw^{*}$ and bias $\rho^{*}$ by solving:
\begin{align*}
\left(G^{*},\bw^{*},\rho^{*}\right) & =\argmin{G,\bw,\rho}\mathcal{I}\left(G,D,\bw,\rho\right)\\
D^{*} & =\argmax D\,\,\mathcal{I}\left(G,D,\bw,\rho\right)
\end{align*}

\section{Experiments\label{sec:ex}}

\paragraph{Experimental Datasets}

We used the same real-world multimedia and image application  datasets as those studied in \citet{vannguyen2019dan,van-nguyen-dual-dan-2020}.
These contain the source code of vulnerable functions (vul-funcs)
and non-vulnerable functions (non-vul-funcs) obtained from six real-world
software project datasets, namely FFmpeg (\#vul-funcs: 187 and \#non-vul-funcs:
5427), LibTIFF (\#vul-funcs: 81 and \#non-vul-funcs: 695), LibPNG
(\#vul-funcs: 43 and \#non-vul-funcs: 551), VLC (\#vul-funcs: 25 and
\#non-vul-funcs: 5548), and Pidgin (\#vul-funcs: 42 and \#non-vul-funcs:
8268). %These datasets cover multimedia and image application categories.

In the experiments, to demonstrate the capability of our proposed
method in  transfer learning for cross-domain software vulnerability detection
(SVD) (i.e., transferring the learning of software vulnerabilities
(SVs) from labelled projects to unlabelled projects belonging to different
application domains), we used the multimedia application datasets (FFmpeg, VLC, and Pidgin) as the source domains,
whilst the datasets (LibPNG and LibTIFF) from the image application
domains were used as the target domains. It is worth noting that in
the training process we hide the labels of datasets from the target
domains. We only use these labels in the testing phase to evaluate
the models\textquoteright{} performance. Moreover, we used 80\% of
the target domain without labels in the training process, while the
rest 20\% was used for evaluating the domain adaptation performance.
Note that these partitions were split randomly as in the baselines.
%\emph{Please refer to the supplementary material for details.}

\paragraph{Baselines \label{subsec:Baselines}}

The main baselines of our proposed DAM2P method are the state-of-the-art
end-to-end deep domain adaptation (DA) approaches for cross-domain SVD including SCDAN \cite{vannguyen2019dan},
Dual-GD-DDAN, Dual-GD-SDDAN \cite{van-nguyen-dual-dan-2020}, DDAN \cite{Ganin2015},
MMD \cite{long2015}, D2GAN \cite{duald-tunguyen2017}, DIRT-T \cite{shu2018a},
HoMM \cite{Hommchen2020}, and LAMDA \cite{LAMDAle21a} as
well as the state-of-the-art automatic feature learning for SVD, VulDeePecker
\cite{VulDeePecker2018}. To the method operated via separated stages
proposed by \citet{Liu2020CD-VulD}, at present, we cannot compare
to it due to the lack of the original data and completed reproducing source
code from the authors.

VulDeePecker \cite{VulDeePecker2018} is an automatic feature learning
method for SVD. The model employed a bidirectional recurrent neural
network to take sequential inputs and then concatenated hidden units
as inputs to a feedforward neural network classifier while the DDAN,
MMD, D2GAN, DIRT-T HoMM, and LAMDA methods are the state-of-the-art
deep domain adaptation models for computer vision proposed in \citet{Ganin2015},
\citet{long2015}, \citet{duald-tunguyen2017}, \citet{shu2018a}, \citet{Hommchen2020},
and \citet{LAMDAle21a} respectively. Inspired by \citet{vannguyen2019dan},
we borrowed the principles of these methods and refactored them using
the CDAN architecture introduced in \citet{vannguyen2019dan} for cross-domain SVD.

The SCDAN method \cite{vannguyen2019dan} can be considered as the
first method that demonstrates the feasibility of deep domain adaptation
for cross-domain SVD. Based on their proposed CDAN architecture, leveraging deep
domain adaptation with automatic feature learning for SVD, the authors
proposed the SCDAN method to efficiently exploit and utilize the information
from unlabeled target domain data to improve the model performance.\textbf{
}The Dual-GD-DDAN and Dual-GD-SDDAN methods were proposed in \citet{van-nguyen-dual-dan-2020}
aiming to deal with the mode collapsing problem in SCDAN and
other approaches (i.e., using GAN as a principle in order to close
the gap between the source and target domains in the joint space)
to further improve the transfer learning process for cross-domain SVD.

\emph{For the data processing and embedding, the models' configuration, the ablation study about the hyper-parameter sensitivity of our DAM2P method, as well as the instructions for reproducing the experimental results,
please refer to the appendix.}

\begin{table}[th]
\vspace{-0.5mm}

\centering{}\resizebox{0.93\columnwidth}{!}{%%
\begin{tabular}{ccccccc}
\hline 
Source $\goto$ Target & Methods & FNR & FPR & Recall & Precision & F1\tabularnewline
\hline 
 & VULD & 42.86\% & 1.08\% & 57.14\% & 80\% & 66.67\%\tabularnewline
 & MMD & 37.50\% & \textbf{0\%} & 62.50\% & \textbf{100\%} & 76.92\%\tabularnewline
 & D2GAN & 33.33\% & 1.06\% & 66.67\% & 80\% & 72.73\%\tabularnewline
 & DIRT-T & 33.33\% & 1.06\% & 66.67\% & 80\% & 72.73\%\tabularnewline
Pidgin\textbf{$\goto$}  & HOMM & 14.29\% & 4.30\% & 85.71\% & 60.00\% & 70.59\%\tabularnewline
~~~LibPNG & LAMDA & \textbf{12.50\%} & 4.35\% & 87.50\% & 63.64\% & 73.68\%\tabularnewline
 & DDAN & 37.50\% & \textbf{0\%} & 62.50\% & \textbf{100\%} & 76.92\%\tabularnewline
 & SCDAN & 33.33\% & \textbf{0\%} & 66.67\% & 100\% & 80\%\tabularnewline
 & Dual-DDAN & 33.33\% & \textbf{0\%} & 66.67\% & \textbf{100\%} & 80\%\tabularnewline
 & Dual-SDDAN & 22.22\% & 1.09\% & 77.78\% & 87.50\% & 82.35\%\tabularnewline
\cline{2-7} \cline{3-7} \cline{4-7} \cline{5-7} \cline{6-7} \cline{7-7} 
 & DAM2P (ours) & \textbf{12.50\%} & 1.08\% & \textbf{87.50\%} & 87.50\% & \textbf{87.50\%}\tabularnewline
\hline 
 & VULD & 43.75\% & 6.72\% & 56.25\% & 50\% & 52.94\%\tabularnewline
 & MMD & 28.57\% & 12.79\% & 71.43\% & 47.62\% & 57.14\%\tabularnewline
 & D2GAN & 30.77\% & 6.97\% & 69.23\% & 64.29\% & 66.67\%\tabularnewline
 & DIRT-T & 25\% & 9.09\% & 75\% & 52.94\% & 62.07\%\tabularnewline
FFmpeg\textbf{$\goto$}  & HOMM & 37.50\% & 2.17\% & 62.50\% & 71.43\% & 66.67\%\tabularnewline
~~~LibTIFF & LAMDA & 37.50\% & \textbf{1.09\%} & 62.50\% & \textbf{88.33\%} & 71.42\%\tabularnewline
 & DDAN & 35.71\% & 6.98\% & 64.29\% & 60\% & 62.07\%\tabularnewline
 & SCDAN & 14.29\% & 5.38\% & 85.71\% & 57.14\% & 68.57\%\tabularnewline
 & Dual-DDAN & \textbf{12.5\%} & 8.2\% & \textbf{87.5\%} & 56\% & 68.29\%\tabularnewline
 & Dual-SDDAN & 35.29\% & 3.01\% & 64.71\% & 73.33\% & 68.75\%\tabularnewline
\cline{2-7} \cline{3-7} \cline{4-7} \cline{5-7} \cline{6-7} \cline{7-7} 
 & DAM2P (ours) & 14.29\% & 8.14\% & 85.71\% & 63.16\% & \textbf{72.73\%}\tabularnewline
\hline 
 & VULD & 25\% & 2.17\% & 75\% & 75\% & 75\%\tabularnewline
 & MMD & 12.5\% & 3.26\% & 87.5\% & 70\% & 77.78\%\tabularnewline
 & D2GAN & 14.29\% & 2.17\% & 85.71\% & 75\% & 80\%\tabularnewline
 & DIRT-T & 15.11\% & 2.2\% & 84.89\% & 80\% & 84.21\%\tabularnewline
FFmpeg\textbf{$\goto$} & HOMM & \textbf{0\%} & 2.15\% & \textbf{100\%} & 77.78\% & 87.50\%\tabularnewline
~~~LibPNG & LAMDA & 16.67\% & \textbf{0.\% } & 83.33\% & \textbf{100\%} & 90.91\%\tabularnewline
 & DDAN & \textbf{0\%} & 3.26\% & \textbf{100\%} & 72.73\% & 84.21\%\tabularnewline
 & SCDAN & 12.5\% & 1.08\% & 87.5\% & 87.5\% & 87.5\%\tabularnewline
 & Dual-DDAN & \textbf{0\%} & 2.17\% & \textbf{100\%} & 80\% & 88.89\%\tabularnewline
 & Dual-SDDAN & 17.5\% & \textbf{0\%} & 82.5\% & \textbf{100\%} & 90.41\%\tabularnewline
\cline{2-7} \cline{3-7} \cline{4-7} \cline{5-7} \cline{6-7} \cline{7-7} 
 & DAM2P (ours) & \textbf{0\%} & 1.07\% & \textbf{100\%} & 87.50\% & \textbf{93.33\%}\tabularnewline
\hline 
 & VULD & 57.14\% & 1.08\% & 42.86\% & 75\% & 54.55\%\tabularnewline
 & MMD & 45\% & 4.35\% & 55\% & 60\% & 66.67\%\tabularnewline
 & D2GAN & 28.57\% & 4.3\% & 71.43\% & 55.56\% & 62.5\%\tabularnewline
 & DIRT-T & 50\% & 1.09\% & 50\% & 80\% & 61.54\%\tabularnewline
VLC\textbf{$\goto$}  & HOMM & 42.86\% & \textbf{0\%} & 57.14\% & \textbf{100\% } & 72.73\%\tabularnewline
~~~~~~LibPNG & LAMDA & 28.57\% & 1.08\% & 71.43\% & 83.33\% & 76.92\%\tabularnewline
 & DDAN & 33.33\% & 2.20\% & 66.67\% & 75\% & 70.59\%\tabularnewline
 & SCDAN & 33.33\% & 1.06\% & 66.67\% & 80\% & 72.73\%\tabularnewline
 & Dual-DDAN & 28.57\% & 2.15\% & 71.43\% & 71.43\% & 71.43\%\tabularnewline
 & Dual-SDDAN & \textbf{11.11\%} & 4.39\% & \textbf{88.89\%} & 66.67\% & 76.19\%\tabularnewline
\cline{2-7} \cline{3-7} \cline{4-7} \cline{5-7} \cline{6-7} \cline{7-7} 
 & DAM2P (ours) & 33.33\% & \textbf{0\%} & 66.67\% & \textbf{100\%} & \textbf{80\%}\tabularnewline
\hline 
 & VULD & 35.29\% & 8.27\% & 64.71\% & 50\% & 56.41\%\tabularnewline
 & MMD & 30.18\% & 12.35\% & 69.82\% & 50\% & 58.27\%\tabularnewline
 & D2GAN & 40\% & 7.95\% & 60\% & 60\% & 60\%\tabularnewline
 & DIRT-T & 38.46\% & 8.05\% & 61.54\% & 53.33\% & 57.14\%\tabularnewline
Pidgin\textbf{$\goto$}  & HOMM & 20\% & 9.41\% & 80\% & 60\% & 68.57\%\tabularnewline
~~~LibTIFF & LAMDA & 30\% & 4.44\% & 70\% & 63.64\% & 66.67\%\tabularnewline
 & DDAN & 27.27\% & 8.99\% & 72.73\% & 50\% & 59.26\%\tabularnewline
 & SCDAN & 30\% & 5.56\% & 70\% & 58.33\% & 63.64\%\tabularnewline
 & Dual-DDAN & 29.41\% & 6.76\% & 70.59\% & 57.14\% & 63.16\%\tabularnewline
 & Dual-SDDAN & 37.5\% & \textbf{2.98\%} & 62.5\% & \textbf{71.43\%} & 66.67\%\tabularnewline
\cline{2-7} \cline{3-7} \cline{4-7} \cline{5-7} \cline{6-7} \cline{7-7} 
 & DAM2P (ours) & \textbf{7.69\%} & 9.20\% & \textbf{92.31\%} & 60\% & \textbf{72.73\%}\tabularnewline
\hline 
\end{tabular}}\vspace{-1mm}
\caption{Performance results in terms of false negative rate (FNR), false positive
rate (FPR), Recall, Precision and F1-measure (F1) of VulDeePecker
(VULD), MMD, D2GAN, DIRT-T, HOMM, LAMDA, DDAN, SCDAN, Dual-GD-DDAN
(Dual-DDAN), Dual-GD-SDDAN (Dual-SDDAN) and DAM2P methods for predicting
vulnerable and non-vulnerable functions on the testing set of the
target domain (Best performance in \textbf{bold}). \label{tab:the-first-case}}
\vspace{-6mm}
\end{table}

\subsection{Experimental Results}

\subsubsection{Domain Adaptation for Non-labeled Target Projects \label{subsec:first_exp}}

\paragraph{Quantitative Results}

We investigated the performance of our DAM2P method and compared
to the baselines. We note that the VulDeePecker method was only
trained on the source domain data and then tested on the target domain data. The
DDAN, MMD, D2GAN, DIRT-T, HOMM, LAMDA, SCDAN, Dual-GD-DDAN, Dual-GD-SDDAN,
and DAM2P methods employed the target domain data without using any label
information for domain adaptation.

The results in Table \ref{tab:the-first-case} show that
our proposed DAM2P method obtains a higher performance for most
measures in the majority of  cases of source and target domains. DAM2P
achieves the highest F1-measure for all pairs of the
source and target domains. In general, our method obtains a higher performance on F1-measure from 1.83\% to 6.25\% compared to the second highest method in the used datasets of the source and target domains. For example, in the case of the source
domain (FFmpeg) and target domain (LibPNG), the DAM2P method obtains
the F1-measure of \emph{93.33\%} compared with the F1-measure of 90.41\%,
88.89\%, 87.5\%, 84.21\%, 90.91\%, 87.50\%, 84.21\%, 80\%, 77.78\%
and 75\% obtained with Dual-GD-SDDAN, Dual-GD-DDAN, SCDAN, DDAN,
LAMDA, HOMM, DIRT-T, D2GAN, MMD and VulDeePecker, respectively.

\paragraph{Visualization}

We further demonstrate the efficiency of our proposed method in closing
the gap between the source and target domains. We visualize the feature
distributions of the source and target domains in the joint space
using a 2D t-SNE \cite{vandermaaten2008visualizing} projection with
perplexity equal to $30$. In particular, we project the source domain and
target domain data in the joint space (i.e., $G\left(\bx\right)$) into a
2D space without undertaking domain adaptation (using the VulDeePecker
method) and with undertaking domain adaptation (using our proposed DAM2P method).

\begin{figure}[th]
\begin{centering}
\vspace{-1mm}
\includegraphics[width=0.92\columnwidth]{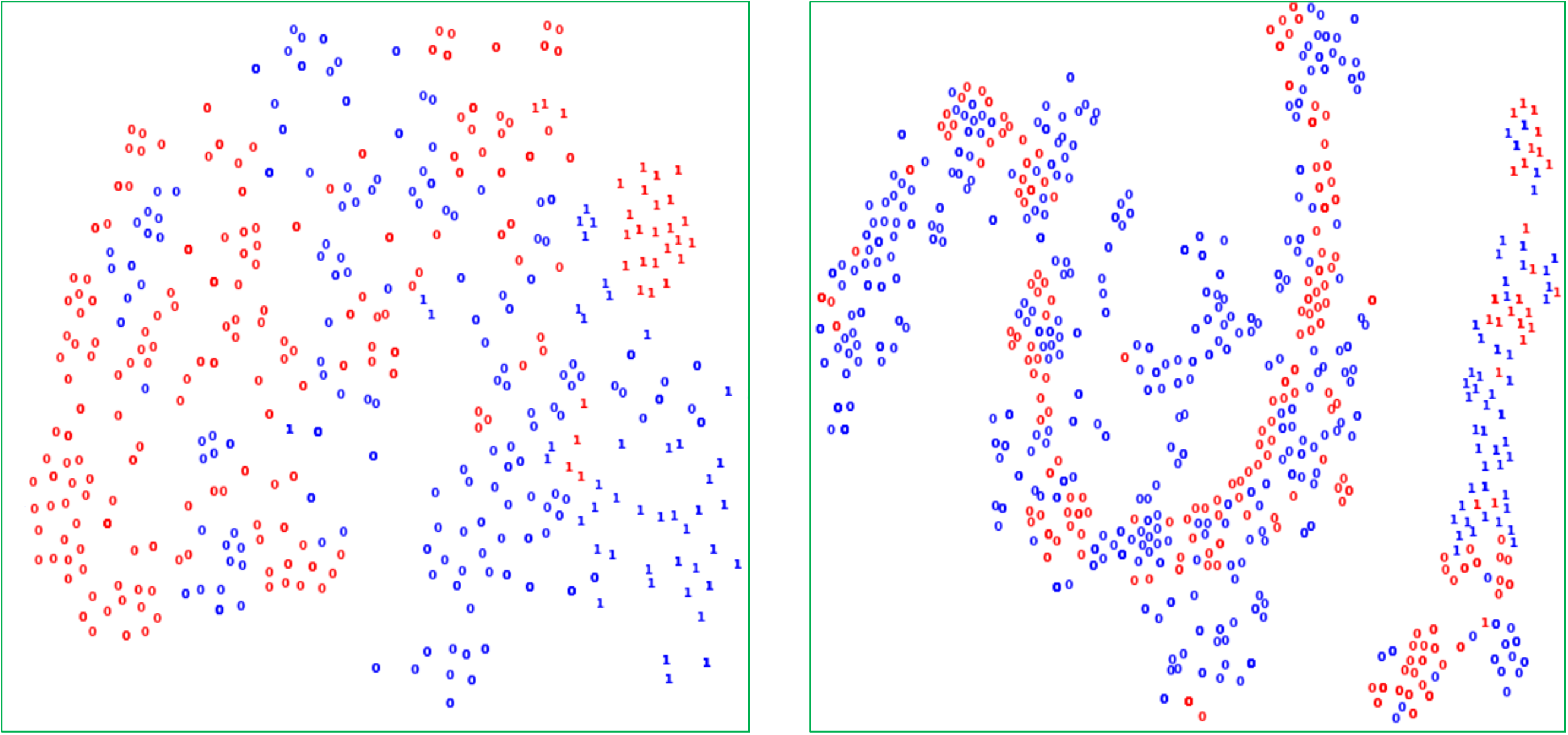}
\par\end{centering}\vspace{-1.5mm}
\caption{A 2D t-SNE projection for the case of the FFmpeg $\protect\goto$
LibPNG without undertaking domain adaptation (the left-hand figure,
using VulDeePecker) and with undertaking domain adaptation
(the right-hand figure, using our proposed DAM2P method). The blue
and red points represent the source and target domains in the joint
space respectively. Data points labeled 0 stand for non-vulnerable
samples and data points labeled 1 stand for vulnerable samples. \emph{It is noted that our method can not only successfully bridge the gap between the source and target domains but also able to distinguish the non-vulnerable and vulnerable data effectively}. \label{fig:visual_peg_png_labels}}
\vspace{-9mm}
\end{figure}

In Figure \ref{fig:visual_peg_png_labels}, we present the results
when performing domain adaptation from a software project (FFmpeg)
to another (LibPNG). For the purpose of visualization, we select a
random subset of the source project against the entire target project.
As shown in Figure \ref{fig:visual_peg_png_labels}, without undertaking
domain adaptation (VulDeePecker) the blue points (the source domain data)
and the red points (the target domain data) are almost separate while with
undertaking domain adaptation the blue and red points intermingled
as expected. Furthermore, we observe that the mixing-up level of source domain
and target domain data using our DAM2P method is significantly higher than
using VulDeePecker.

\paragraph{Ablation Study}

In this section, we aim to further demonstrate the efficiency of our
DAM2P method in transferring the learning of software vulnerabilities
from imbalanced labeled source domains to other imbalanced unlabeled
target domains as well as the superiority of our novel cross-domain
kernel classifier in our DAM2P method for learning and separating
vulnerable and non-vulnerable data.

\begin{table}[h]
\vspace{-1.5mm}
\centering{}\resizebox{0.93\columnwidth}{!}{%%
\begin{tabular}{ccccccc}
\hline 
Source $\goto$ Target & Methods & FNR & FPR & Recall & Precision & F1\tabularnewline
\hline 
 & VulDeePecker & 43.75\% & 6.72\% & 56.25\% & 50\% & 52.94\%\tabularnewline
FFmpeg \textbf{$\goto$}  & DDAN & 35.71\% & 6.98\% & 64.29\% & 60\% & 62.07\%\tabularnewline
~~~LibTIFF & Kernel-S & 30\% & \textbf{5.56\%} & 70\% & 58.33\% & 63.63\%\tabularnewline
 & Kernel-ST & 25\% & 5.68\% & 75\% & \textbf{64.29\%} & 69.23\%\tabularnewline
\cline{2-7} \cline{3-7} \cline{4-7} \cline{5-7} \cline{6-7} \cline{7-7} 
 & DAM2P (ours) & \textbf{14.29\%} & 8.14\% & \textbf{85.71\%} & 63.16\% & \textbf{72.73\%}\tabularnewline
\hline 
 & VulDeePecker & 25\% & 2.17\% & 75\% & 75\% & 75\%\tabularnewline
FFmpeg \textbf{$\goto$}  & DDAN & \textbf{0\%} & 3.26\% & \textbf{100\%} & 72.73\% & 84.21\%\tabularnewline
~~~LibPNG & Kernel-S & \textbf{0\%} & 4.39\% & \textbf{100\%} & 69.23\% & 81.81\%\tabularnewline
 & Kernel-ST & \textbf{0\%} & 3.26\% & \textbf{100\%} & 72.72\% & 84.21\%\tabularnewline
\cline{2-7} \cline{3-7} \cline{4-7} \cline{5-7} \cline{6-7} \cline{7-7} 
 & DAM2P (ours) & \textbf{0\%} & \textbf{1.07\%} & \textbf{100\%} & \textbf{87.50\%} & \textbf{93.33\%}\tabularnewline
\hline 
\end{tabular}}\vspace{-1.5mm}
\caption{Performance results in terms of false negative rate (FNR), false positive
rate (FPR), Recall, Precision and F1-measure (F1) of five cases including
(i, VulDeePecker), (ii, DDAN), (iii, Kernel-Source denoted by Kernel-S),
(iv, Kernel-Source-Target denoted by Kernel-ST), and (v, DAM2P) for
predicting vulnerable and non-vulnerable code functions on the testing
set of the target domain (Best performance in \textbf{bold}). \label{tab:the-first-case-ablation-study}}
\vspace{-3mm}
\end{table}

For this ablation study, we conduct experiments on two pairs FFmpeg
\textbf{$\goto$} LibTIFF and FFmpeg \textbf{$\goto$} LibPNG. We
want to demonstrate that bridging the discrepancy gap in the latent
space and the max-margin cross-domain kernel classifier are complementary
to boost the domain adaptation performance with imbalanced nature.
We consider five cases in which we start from the \emph{blank case
(i, VulDeePecker)} without bridging the gap and cross-domain kernel
classifier. We then only add the GAN term to bridge the discrepancy
gap in the \emph{second case (ii, DDAN)}. In the \emph{third case
(iii, Kernel-Source)}, we only apply the max-margin principle for
the source domain, while applying the max-margin principle for the
source and target domains in the \emph{fourth case (iv, Kernel-Source-Target)}.
Finally, in the \emph{last case (v, DAM2P)}, we simultaneously apply
the bridging term and the max-margin terms for the source and target
domains. The results in Table \ref{tab:the-first-case-ablation-study}
shows that the max-margin terms and bridging term help to boost the
domain adaptation performance. Moreover, applying the max-margin term
to both the source and target domains improves the performance comparing
to applying to only source domain. Last but not least, bridging the
discrepancy gap term in cooperation with the max-margin term significantly
improves the domain adaptation performance.

\paragraph{Additional ablation study}

\paragraph{Sampling and weighting, and Logit adjustment}

Sampling and weighting are well-known as simple and heuristic methods
to deal with imbalanced datasets. However, as mentioned by \citet{Lin2017focal_loss,Cui2019_classbalanced_loss},
these methods may have some limitations, for example, i) Sampling may
either introduce large amounts of duplicated samples, which slows
down the training and makes the model susceptible to overfitting when
oversampling, or discard valuable examples that are important for
feature learning when undersampling, and ii) To the highly imbalanced
datasets, directly training the model or weighting (e.g., inverse
class frequency or the inverse square root of class frequency) cannot
yield satisfactory performance.

\citet{la2021} recently proposed a novel statistical framework for
solving the imbalanced (long-tailed) label distribution problem. Specifically,
the framework, revisiting the idea of \emph{logit adjustment} based
on the label frequencies, encourages a large relative margin between
logits of the rare positive labels versus the dominant negative labels.

To experience and investigate the efficiency of these methods (i.e.,
\emph{sampling and weighting, and logit adjustment}) when applying
to the baselines in the context of imbalanced domain adaptation, we
conduct an experiment on two pairs of the source and target domains
(i.e., FFmpeg \textbf{$\goto$} LibTIFF and FFmpeg \textbf{$\goto$}
LibPNG) for four main baselines including DDAN, SCDAN, Dual-GD-DDAN,
and Dual-GD-SDDAN using (i) the oversampling technique based on SMOTE
\cite{Chawla2002_smote} (i.e., used to create balanced datasets),
and (ii) \emph{logit adjustment} (LA) used in \citet{la2021}.

The results in Table \ref{tab:the-over-sampling} show that using
the oversampling technique cannot help to improve these baseline models\textquoteright{}
performance. In particular, the performance of these baselines without
using oversampling is always higher than using oversampling on the
used datasets in F1-measure (F1), the most important measure used
in SVD. This experiment supports our
conjecture that in the context of imbalanced domain adaptation when
moving target representations to source representations in the latent
space to bridge the gap, oversampling the minority class (i.e., vulnerable
class) might increase the chance to wrongly mix up vulnerable representations
of the target domain and non-vulnerable representations of the source
domain, hence leading to a reduction in performance. In our approach,
with the support of the max-margin principle, we keep the vulnerable
and non-vulnerable representations distant as much as possible when
bridging the gap between them in the latent space.

\begin{table}[th]
\vspace{-1mm}

\centering{}\resizebox{0.93\columnwidth}{!}{%%
\begin{tabular}{clccccc}
\hline 
S $\goto$ T & Methods & FNR & FPR & Recall & Precision & F1\tabularnewline
\hline 
 & DDAN w/ OS & 21\% & 12.79\% & 78.57\% & 50\% & 61.11\%\tabularnewline
 & DDAN w/ LA & 14.29\% & 15.11\% & 85.71\% & 48\% & 61.53\%\tabularnewline
 & DDAN w/o (OS or LA) & 35.71\% & 6.98\% & 64.29\% & 60\% & \textbf{62.07\%}\tabularnewline
\cline{2-7} \cline{3-7} \cline{4-7} \cline{5-7} \cline{6-7} \cline{7-7} 
 & SCDAN w/ OS & 25\% & 5.43\% & 75\% & 54.55\% & 63.16\%\tabularnewline
 & SCDAN w/ LA & 11.11\% & 8.8\% & 88.89\% & 50\% & 64\%\tabularnewline
FFmpeg \textbf{$\goto$}  & SCDAN w/o (OS or LA) & 14.29\% & 5.38\% & 85.71\% & 57.14\% & \textbf{68.57\%}\tabularnewline
\cline{2-7} \cline{3-7} \cline{4-7} \cline{5-7} \cline{6-7} \cline{7-7} 
~~~LibTIFF & Dual-DDAN w/ OS & 25\% & 6.72\% & 75\% & 57.14\% & 64.87\%\tabularnewline
 & Dual-DDAN w/ LA & 35.29\% & 4.50\% & 64.71\% & 64.71\% & 64.71\%\tabularnewline
 & Dual-DDAN w/o (OS or LA) & 12.5\% & 8.2\% & 87.5\% & 56\% & \textbf{68.29\%}\tabularnewline
\cline{2-7} \cline{3-7} \cline{4-7} \cline{5-7} \cline{6-7} \cline{7-7} 
 & Dual-SDDAN w/ OS & 16.67\% & 9.1\% & 83.33\% & 56\% & 67\%\tabularnewline
 & Dual-SDDAN w/ LA & 43.75\% & 1.70\% & 56.25\% & 90\% & \textbf{69\%}\tabularnewline
 & Dual-SDDAN w/o (OS or LA) & 35.29\% & 3.01\% & 64.71\% & 73.33\% & 68.75\%\tabularnewline
\hline 
 & DDAN w/ OS & 28.57\% & 0\% & 71.43\% & 100\% & 83.33\% \tabularnewline
 & DDAN w/ LA & 25\% & 0\% & 75\% & 100\% & \textbf{85.71\%}\tabularnewline
 & DDAN w/o (OS or LA) & 0\% & 3.26\% & 100\% & 72.73\% & 84.21\%\tabularnewline
\cline{2-7} \cline{3-7} \cline{4-7} \cline{5-7} \cline{6-7} \cline{7-7} 
 & SCDAN w/ OS & 25\% & 0\% & 75\% & 100\% & 85.71\%\tabularnewline
 & SCDAN w/ LA & 0\% & 2.17\% & 100\% & 80\% & \textbf{88.89\%}\tabularnewline
FFmpeg \textbf{$\goto$}  & SCDAN w/o (OS or LA) & 12.5\% & 1.08\% & 87.5\% & 87.5\% & 87.5\%\tabularnewline
\cline{2-7} \cline{3-7} \cline{4-7} \cline{5-7} \cline{6-7} \cline{7-7} 
~~~LibPNG & Dual-DDAN w/ OS & 14.29\% & 1.08\% & 85.71\% & 85.71\% & 85.71\%\tabularnewline
 & Dual-DDAN w/ LA & 0\% & 2.15\% & 100\% & 77.78\% & 87.5\%\tabularnewline
 & Dual-DDAN w/o (OS or LA) & 0\% & 2.17\% & 100\% & 80\% & \textbf{88.89\%}\tabularnewline
\cline{2-7} \cline{3-7} \cline{4-7} \cline{5-7} \cline{6-7} \cline{7-7} 
 & Dual-SDDAN w/ OS & 12.5\% & 2.17\% & 87.5\% & 77.78\% & 82.35\%\tabularnewline
 & Dual-SDDAN w/ LA & 11.11\% & 1.1\% & 88.89\% & 88.89\% & 88.89\%\tabularnewline
 & Dual-SDDAN w/o (OS or LA) & 17.5\% & 0\% & 82.5\% & 100\% & \textbf{90.41\%}\tabularnewline
\hline 
\end{tabular}}\vspace{-1mm}
\caption{Performance results in terms of false negative rate (FNR), false positive
rate (FPR), Recall, Precision, and F1-measure (F1) of DDAN, SCDAN,
Dual-GD-DDAN (Dual-DDAN), and Dual-GD-SDDAN (Dual-SDDAN) methods in
three cases of with using oversampling (w/ OS), using LA (w/ LA) and
without using (oversampling or LA) (w/o (OS or LA)) for predicting
vulnerable and non-vulnerable code functions on the testing set of
the target domain. We denote Source $\protect\goto$ Target by S $\protect\goto$
T. \label{tab:the-over-sampling}}
\vspace{-2mm}
\end{table}

Furthermore, the results in Table \ref{tab:the-over-sampling} indicate
that using LA can help slightly improve the model's performance on
some cases of the baselines. In particular, LA increases the performance
of Dual-SDDAN on FFmpeg \textbf{$\goto$} LibTIFF as well as DDAN
and SCDAN on FFmpeg \textbf{$\goto$} LibPNG compared to these methods
without using LA. However, to DDAN, SCDAN and Dual-DDAN on FFmpeg
\textbf{$\goto$} LibTIFF as well as Dual-DDAN and Dual-SDDAN on FFmpeg
\textbf{$\goto$} LibPNG, LA cannot help improve these models' performance.
In conclusion, using LA can help increase the baseline's performance
in some cases of the used datasets in F1-measure compared to these
cases without using LA or using the oversampling technique. However,
similar to the oversampling technique, in most cases mentioned in
Table \ref{tab:the-over-sampling}, LA cannot help improve the baselines'
performance. Furthermore, in the cases where LA helps increase the
baseline models' performance, our proposed method's performance (mentioned
in Table \ref{tab:the-first-case}) is still significantly higher.

%\emph{Please refer to the supplementary material for the ablationstudy about the hyper-parameter sensitivity of our proposed DAM2Pmethod.}

\section{Conclusion\label{sec:conclusion}}

In this paper, in addition to exploiting deep domain adaptation with
automatic representation learning for SVD, we have successfully proposed
a novel cross-domain kernel classifier leveraging the max-margin principle
to significantly improve the capability of the transfer learning of
software vulnerabilities from labeled projects into unlabeled ones
in order to deal with two crucial issues in SVD including i) learning
automatic representations to improve the predictive performance of
SVD, and ii) coping with the scarcity of labeled vulnerabilities in
projects that require the laborious labeling of code by experts. Our
proposed cross-domain kernel classifier can not only effectively deal
with the imbalanced datasets but also leverage the information of
the unlabeled projects to further improve the classifier's performance.
The experimental results show the superiority of our proposed method
compared with other state-of-the-art baselines in terms of the representation
learning and transfer learning processes.

%\newpage
%\bibliographystyle{aaai23}
\bibliography{sigproc}

\begin{thebibliography}{42}
\providecommand{\natexlab}[1]{#1}

\bibitem[{Abadi et~al.(2016)Abadi, Barham, Chen, Chen, Davis, Dean, Devin,
  Ghemawat, Irving, Isard et~al.}]{abadi2016tensorflow}
Abadi, M.; Barham, P.; Chen, J.; Chen, Z.; Davis, A.; Dean, J.; Devin, M.;
  Ghemawat, S.; Irving, G.; Isard, M.; et~al. 2016.
\newblock Tensorflow: A system for large-scale machine learning.
\newblock In \emph{12th USENIX Symposium on Operating Systems Design and
  Implementation (OSDI 16)}, 265--283.

\bibitem[{Chawla et~al.(2002)Chawla, Bowyer, Hall, and
  Kegelmeyer}]{Chawla2002_smote}
Chawla, N.~V.; Bowyer, K.~W.; Hall, L.~O.; and Kegelmeyer, W.~P. 2002.
\newblock SMOTE: synthetic minority over-sampling technique.
\newblock \emph{Journal of Artificial Intelligence Research}, 16: 321--357.

\bibitem[{Chen et~al.(2020)Chen, Fu, Chen, Jin, Cheng, Jin, and
  Hua}]{Hommchen2020}
Chen, C.; Fu, Z.; Chen, Z.; Jin, S.; Cheng, Z.; Jin, X.; and Hua, X. 2020.
\newblock HoMM: Higher-order Moment Matching for Unsupervised Domain
  Adaptation.
\newblock \emph{Thirty-Fourth AAAI Conference on Artificial Intelligence}.

\bibitem[{Cheng et~al.(2019)Cheng, Wang, Hua, Zhang, Xu, Yi, and
  Sui}]{Cheng2019}
Cheng, X.; Wang, H.; Hua, J.; Zhang, M.; Xu, G.; Yi, L.; and Sui, Y. 2019.
\newblock Static Detection of Control-Flow-Related Vulnerabilities Using Graph
  Embedding.
\newblock In \emph{2019 24th International Conference on Engineering of Complex
  Computer Systems (ICECCS)}.

\bibitem[{Cui et~al.(2019)Cui, Jia, Lin, Song, and
  Belongie}]{Cui2019_classbalanced_loss}
Cui, Y.; Jia, M.; Lin, T.; Song, Y.; and Belongie, S.~J. 2019.
\newblock Class-Balanced Loss Based on Effective Number of Samples.
\newblock \emph{CoRR}, abs/1901.05555.

\bibitem[{Dam et~al.(2018)Dam, Tran, Pham, Wee, Grundy, and Ghose}]{Dam2018}
Dam, H.~K.; Tran, T.; Pham, T.; Wee, N.~S.; Grundy, J.; and Ghose, A. 2018.
\newblock Automatic feature learning for predicting vulnerable software
  components.
\newblock \emph{IEEE Transactions on Software Engineering}.

\bibitem[{Dowd, McDonald, and Schuh(2006)}]{Dowd2006}
Dowd, M.; McDonald, J.; and Schuh, J. 2006.
\newblock \emph{The Art of Software Security Assessment: Identifying and
  Preventing Software Vulnerabilities}.
\newblock Addison-Wesley Professional.
\newblock ISBN 0321444426.

\bibitem[{Duan et~al.(2019)Duan, Wu, Ji, Rui, Luo, Yang, and Wu}]{Duan2019}
Duan, X.; Wu, J.; Ji, S.; Rui, Z.; Luo, T.; Yang, M.; and Wu, Y. 2019.
\newblock VulSniper: Focus Your Attention to Shoot Fine-Grained
  Vulnerabilities.
\newblock In \emph{Proceedings of the Twenty-Eighth International Joint
  Conference on Artificial Intelligence, {IJCAI-19}}.

\bibitem[{Ganin and Lempitsky(2015)}]{Ganin2015}
Ganin, Y.; and Lempitsky, V. 2015.
\newblock Unsupervised Domain Adaptation by Backpropagation.
\newblock In \emph{Proceedings of the 32nd International Conference on
  International Conference on Machine Learning - Volume 37}, ICML'15,
  1180--1189.

\bibitem[{Goodfellow et~al.(2014)Goodfellow, Pouget-Abadie, Mirza, Xu,
  Warde-Farley, Ozair, Courville, and Bengio}]{goodfellow2014generative}
Goodfellow, I.; Pouget-Abadie, J.; Mirza, M.; Xu, B.; Warde-Farley, D.; Ozair,
  S.; Courville, A.; and Bengio, Y. 2014.
\newblock Generative adversarial nets.
\newblock In \emph{Advances in neural information processing systems},
  2672--2680.

\bibitem[{Grieco et~al.(2016)Grieco, Grinblat, Uzal, Rawat, Feist, and
  Mounier}]{Grieco2016}
Grieco, G.; Grinblat, G.~L.; Uzal, L.; Rawat, S.; Feist, J.; and Mounier, L.
  2016.
\newblock Toward Large-Scale Vulnerability Discovery Using Machine Learning.
\newblock In \emph{Proceedings of the Sixth ACM Conference on Data and
  Application Security and Privacy}, CODASPY '16, 85--96.
\newblock ISBN 978-1-4503-3935-3.

\bibitem[{Hochreiter and Schmidhuber(1997)}]{HochSchm97}
Hochreiter, S.; and Schmidhuber, J. 1997.
\newblock Long Short-Term Memory.
\newblock \emph{Neural Computation}, 9(8): 1735--1780.

\bibitem[{Kim et~al.(2017)Kim, Woo, Lee, and Oh}]{KimWLO17}
Kim, S.; Woo, S.; Lee, H.; and Oh, H. 2017.
\newblock {VUDDY:} {A} Scalable Approach for Vulnerable Code Clone Discovery.
\newblock In \emph{{IEEE} Symposium on Security and Privacy}, 595--614. {IEEE}
  Computer Society.

\bibitem[{Kingma and Ba(2014)}]{KingmaB14}
Kingma, D.~P.; and Ba, J. 2014.
\newblock Adam: {A} Method for Stochastic Optimization.
\newblock \emph{CoRR}, abs/1412.6980.

\bibitem[{Kipf and Welling(2016)}]{KipfW16}
Kipf, T.~N.; and Welling, M. 2016.
\newblock Semi-Supervised Classification with Graph Convolutional Networks.
\newblock \emph{CoRR}, abs/1609.02907.

\bibitem[{Laurens and Geoffrey(2008)}]{vandermaaten2008visualizing}
Laurens, V.~M.; and Geoffrey, H. 2008.
\newblock Visualizing Data using {t-SNE}.
\newblock \emph{Journal of Machine Learning Research}.

\bibitem[{Le et~al.(2021)Le, Nguyen, Ho, Bui, and Phung}]{LAMDAle21a}
Le, T.; Nguyen, T.; Ho, N.; Bui, H.; and Phung, D. 2021.
\newblock LAMDA: Label Matching Deep Domain Adaptation.
\newblock In \emph{Proceedings of the 38th International Conference on Machine
  Learning}, 6043--6054.

\bibitem[{Le et~al.(2010)Le, Tran, Ma, and Sharma}]{trung2010}
Le, T.; Tran, D.; Ma, W.; and Sharma, D. 2010.
\newblock An optimal sphere and two large margins approach for novelty
  detection.
\newblock In \emph{Neural Networks (IJCNN), The 2010 International Joint
  Conference on}, 1--6.

\bibitem[{Li et~al.(2016)Li, Zou, Xu, Jin, Qi, and Hu}]{Li2016:VAV}
Li, Z.; Zou, D.; Xu, S.; Jin, H.; Qi, H.; and Hu, J. 2016.
\newblock VulPecker: An Automated Vulnerability Detection System Based on Code
  Similarity Analysis.
\newblock In \emph{Proceedings of the 32Nd Annual Conference on Computer
  Security Applications}, ACSAC '16, 201--213.
\newblock ISBN 978-1-4503-4771-6.

\bibitem[{Li et~al.(2018{\natexlab{a}})Li, Zou, Xu, Jin, Zhu, Chen, Wang, and
  Wang}]{Li2018SySeVR}
Li, Z.; Zou, D.; Xu, S.; Jin, H.; Zhu, Y.; Chen, Z.; Wang, S.; and Wang, J.
  2018{\natexlab{a}}.
\newblock SySeVR: {A} Framework for Using Deep Learning to Detect Software
  Vulnerabilities.
\newblock \emph{CoRR}, abs/1807.06756.

\bibitem[{Li et~al.(2018{\natexlab{b}})Li, Zou, Xu, Ou, Jin, Wang, Deng, and
  Zhong}]{VulDeePecker2018}
Li, Z.; Zou, D.; Xu, S.; Ou, X.; Jin, H.; Wang, S.; Deng, Z.; and Zhong, Y.
  2018{\natexlab{b}}.
\newblock VulDeePecker: {A} Deep Learning-Based System for Vulnerability
  Detection.
\newblock \emph{CoRR}, abs/1801.01681.

\bibitem[{Lin et~al.(2018)Lin, Zhang, Luo, Pan, Xiang, Olivier, and
  Paul}]{jun_2018}
Lin, G.; Zhang, J.; Luo, W.; Pan, L.; Xiang, Y.; Olivier, D.~V.; and Paul, M.
  2018.
\newblock Cross-Project Transfer Representation Learning for Vulnerable
  Function Discovery.
\newblock In \emph{IEEE Transactions on Industrial Informatics}, volume~14.

\bibitem[{Lin et~al.(2017)Lin, Goyal, Girshick, He, and
  Doll{\'{a}}r}]{Lin2017focal_loss}
Lin, T.; Goyal, P.; Girshick, R.~B.; He, K.; and Doll{\'{a}}r, P. 2017.
\newblock Focal Loss for Dense Object Detection.
\newblock \emph{CoRR}, abs/1708.02002.

\bibitem[{Liu et~al.(2020)Liu, Lin, Qu, Zhang, De~Vel, Montague, and
  Xiang}]{Liu2020CD-VulD}
Liu, S.; Lin, G.; Qu, L.; Zhang, J.; De~Vel, O.; Montague, P.; and Xiang, Y.
  2020.
\newblock CD-VulD: Cross-Domain Vulnerability Discovery based on Deep Domain
  Adaptation.
\newblock \emph{IEEE Transactions on Dependable and Secure Computing}.

\bibitem[{Long et~al.(2015)Long, Cao, Wang, and Jordan}]{long2015}
Long, M.; Cao, Y.; Wang, J.; and Jordan, M. 2015.
\newblock Learning Transferable Features with Deep Adaptation Networks.
\newblock In \emph{Proceedings of the 32nd International Conference on Machine
  Learning}, 97--105.

\bibitem[{Menon et~al.(2021)Menon, Jayasumana, Rawat, Jain, Veit, and
  Kumar}]{la2021}
Menon, A.~K.; Jayasumana, S.; Rawat, A.~S.; Jain, H.; Veit, A.; and Kumar, S.
  2021.
\newblock Long-tail learning via logit adjustment.
\newblock \emph{International Conference on Learning Representations}.

\bibitem[{Neuhaus et~al.(2007)Neuhaus, Zimmermann, Holler, and
  Zeller}]{Neuhaus:2007:PVS}
Neuhaus, S.; Zimmermann, T.; Holler, C.; and Zeller, A. 2007.
\newblock Predicting Vulnerable Software Components.
\newblock In \emph{Proceedings of the 14th ACM Conference on Computer and
  Communications Security}, CCS '07, 529--540.
\newblock ISBN 978-1-59593-703-2.

\bibitem[{Nguyen et~al.(2017)Nguyen, Le, Vu, and Phung}]{duald-tunguyen2017}
Nguyen, T.~D.; Le, T.; Vu, H.; and Phung, D.~Q. 2017.
\newblock Dual Discriminator Generative Adversarial Nets.
\newblock \emph{CoRR}, abs/1709.03831.

\bibitem[{Nguyen et~al.(2020)Nguyen, Le, De~Vel, Montague, Grundy, and
  Phung}]{van-nguyen-dual-dan-2020}
Nguyen, V.; Le, T.; De~Vel, O.; Montague, P.; Grundy, J.; and Phung, D. 2020.
\newblock Dual-Component Deep Domain Adaptation: A New Approach for Cross
  Project Software Vulnerability Detection.

\bibitem[{Nguyen et~al.(2019)Nguyen, Le, Le, Nguyen, DeVel, Montague, Qu, and
  Phung}]{vannguyen2019dan}
Nguyen, V.; Le, T.; Le, T.; Nguyen, K.; DeVel, O.; Montague, P.; Qu, L.; and
  Phung, D. 2019.
\newblock Deep Domain Adaptation for Vulnerable Code Function Identification.
\newblock In \emph{The International Joint Conference on Neural Networks
  (IJCNN)}.

\bibitem[{Nguyen et~al.(2014)Nguyen, Le, Pham, Dinh, and Le}]{Van2014}
Nguyen, V.; Le, T.; Pham, T.; Dinh, M.; and Le, T.~H. 2014.
\newblock Kernel-based semi-supervised learning for novelty detection.
\newblock In \emph{2014 International Joint Conference on Neural Networks
  (IJCNN)}, 4129--4136.

\bibitem[{Rahimi and Recht(2008)}]{Rahimi_NIPS2007}
Rahimi, A.; and Recht, B. 2008.
\newblock Random Features for Large-Scale Kernel Machines.
\newblock In \emph{Advances in Neural Information Processing Systems}.

\bibitem[{Russell et~al.(2018)Russell, Kim, Hamilton, Lazovich, Harer, Ozdemir,
  Ellingwood, and McConley}]{Rebecca2018}
Russell, R.~L.; Kim, L.~Y.; Hamilton, L.~H.; Lazovich, T.; Harer, J.~A.;
  Ozdemir, O.; Ellingwood, P.~M.; and McConley, M.~W. 2018.
\newblock Automated Vulnerability Detection in Source Code Using Deep
  Representation Learning.
\newblock \emph{CoRR}, abs/1807.04320.

\bibitem[{Sch\"{o}lkopf et~al.(2001)Sch\"{o}lkopf, Platt, Shawe-Taylor, Smola,
  and Williamson}]{scholkopf2001}
Sch\"{o}lkopf, B.; Platt, J.~C.; Shawe-Taylor, J.~C.; Smola, A.~J.; and
  Williamson, R.~C. 2001.
\newblock Estimating the Support of a High-Dimensional Distribution.
\newblock \emph{Neural Comput.}, 13(7): 1443--1471.

\bibitem[{Shin et~al.(2011)Shin, Meneely, Williams, and
  Osborne}]{shin2011evaluating}
Shin, Y.; Meneely, A.; Williams, L.; and Osborne, J.~A. 2011.
\newblock Evaluating complexity, code churn, and developer activity metrics as
  indicators of software vulnerabilities.
\newblock \emph{IEEE Transactions on Software Engineering}, 37(6): 772--787.

\bibitem[{Shu et~al.(2018)Shu, Bui, Narui, and Ermon}]{shu2018a}
Shu, R.; Bui, H.; Narui, H.; and Ermon, S. 2018.
\newblock A {DIRT}-T Approach to Unsupervised Domain Adaptation.
\newblock In \emph{International Conference on Learning Representations}.

\bibitem[{Tax and Duin(2004)}]{tax_svdd}
Tax, D. M.~J.; and Duin, R. P.~W. 2004.
\newblock Support Vector Data Description.
\newblock \emph{Journal of Machine Learning Research}, 54(1): 45--66.

\bibitem[{Tsang, Kocsor, and Kwok(2007)}]{Tsang2007}
Tsang, I.~W.; Kocsor, A.; and Kwok, J.~T. 2007.
\newblock Simpler Core Vector Machines with Enclosing Balls.
\newblock In \emph{Proceedings of the 24th International Conference on Machine
  Learning}, ICML '07, 911--918.

\bibitem[{Tsang et~al.(2005)Tsang, Kwok, Cheung, and
  Cristianini}]{Tsang05corevector}
Tsang, I.~W.; Kwok, J.~T.; Cheung, P.; and Cristianini, N. 2005.
\newblock Core vector machines: Fast SVM training on very large data sets.
\newblock \emph{Journal of Machine Learning Research}, 6: 363--392.

\bibitem[{Yamaguchi, Lindner, and Rieck(2011)}]{yamaguchi2011vulnerability}
Yamaguchi, F.; Lindner, F.; and Rieck, K. 2011.
\newblock Vulnerability extrapolation: assisted discovery of vulnerabilities
  using machine learning.
\newblock In \emph{Proceedings of the 5th USENIX conference on Offensive
  technologies}, 13--23.

\bibitem[{Zhuang et~al.(2020)Zhuang, Liu, Qian, Liu, Wang, and He}]{Zhuang2020}
Zhuang, Y.; Liu, Z.; Qian, P.; Liu, Q.; Wang, X.; and He, Q. 2020.
\newblock Smart Contract Vulnerability Detection using Graph Neural Network.
\newblock In \emph{Proceedings of the Twenty-Ninth International Joint
  Conference on Artificial Intelligence, {IJCAI-20}}.

\bibitem[{Zimmermann et~al.(2009)Zimmermann, Nagappan, Gall, Giger, and
  Murphy}]{Zimmermann2009}
Zimmermann, T.; Nagappan, N.; Gall, H.; Giger, E.; and Murphy, B. 2009.
\newblock Cross-project Defect Prediction: A Large Scale Experiment on Data vs.
  Domain vs. Process.
\newblock In \emph{Proceedings of the the 7th Joint Meeting of the European
  Software Engineering Conference and the ACM SIGSOFT Symposium on The
  Foundations of Software Engineering}, ESEC/FSE '09, 91--100.
\newblock ISBN 978-1-60558-001-2.

\end{thebibliography}

\newpage
\appendix
\section{Appendix}
\vspace{1mm}

\subsection{Data Processing and Embedding}

We preprocess the source code datasets before inputting them into the deep neural
networks (i.e., baselines and our proposed method). Inspired from
the baselines, we first standardize
the source code by removing comments, blank lines and non-ASCII characters.
Secondly, we map user-defined variables to symbolic variable names (e.g., \textquotedblleft \emph{var1}\textquotedblright ,
\textquotedblleft \emph{var2}\textquotedblright ) and user-defined
functions to symbolic function names (e.g., \textquotedblleft \emph{func1}\textquotedblright ,
\textquotedblleft \emph{func2}\textquotedblright ). We also replace integer,
real and hexadecimal numbers with a generic <\emph{number}> token
and strings with a generic <\emph{str}> token. We then embed the source code statements into numeric vectors. For example, to the following code statement \emph{``if(func2(func3(number,number),\&var2)
!=var10)}'', we tokenize
it to a sequence of code tokens (e.g., \emph{if,(,func2,(,func3,(,number,number,),\&,var2,),!=,var10,)}),
construct the frequency vector of the statement information, and multiply
this frequency vector by a learnable embedding matrix $W^{si}$.

\begin{figure}[H]
\begin{centering}
\includegraphics[width=0.95\columnwidth]{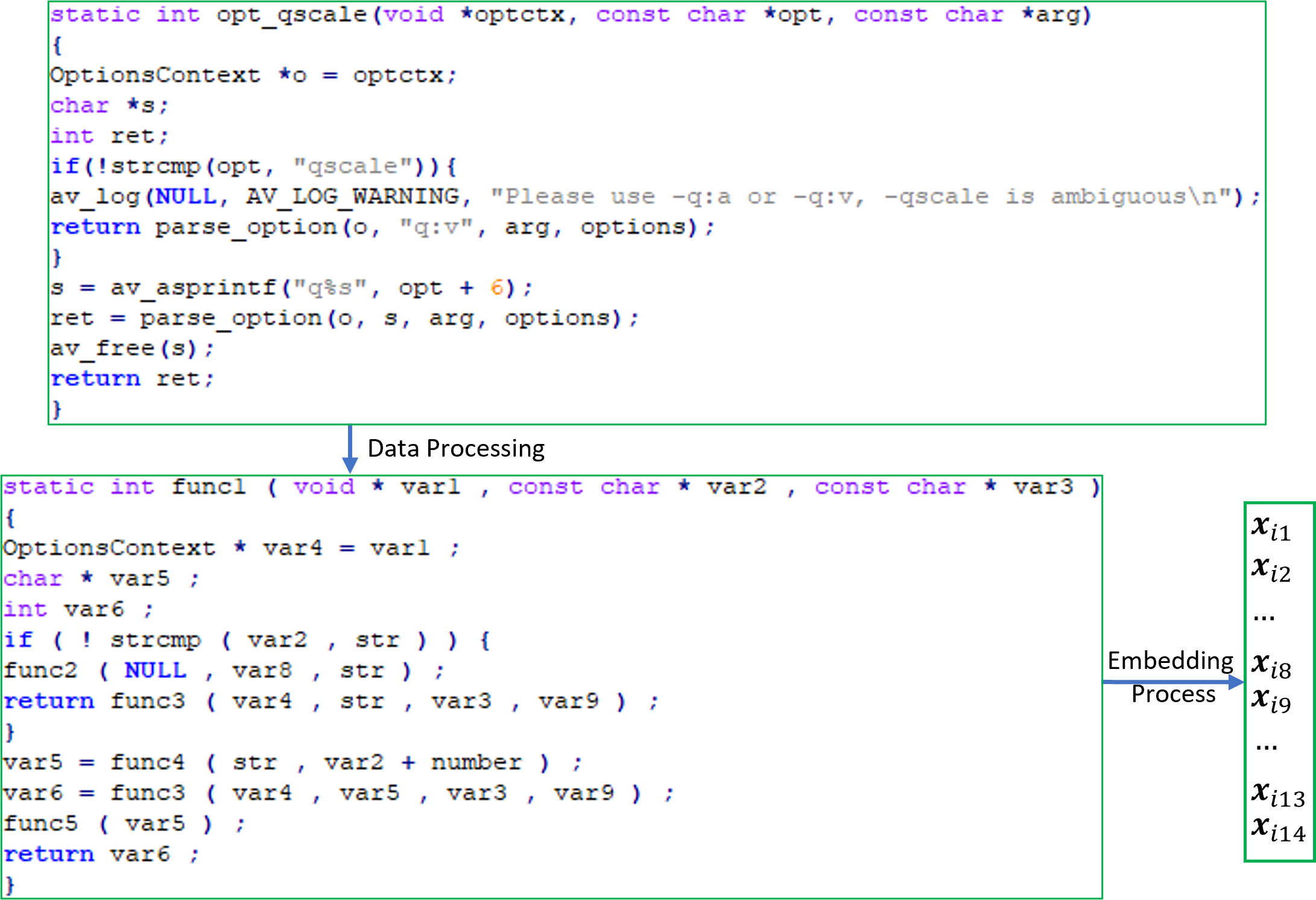}
\par\end{centering}
\caption{An example of the overall procedure for data processing and embedding. We
use a source code function in the C language programming from the
FFmpeg project. After the data preprocessing step, we obtain a preprocessed
function, and then using the embedding process to obtain the embedded
vectors for the code statements of the function.\label{fig:An-example-of-processed-data}}
\vspace{-1mm}
\end{figure}

Figure \ref{fig:An-example-of-processed-data} shows an example of
the overall procedure for a source code function processing and embedding. The sequence of $L$ embedding vectors
(e.g., $\bx_{i}=\left[\boldsymbol{x}_{i1},\dots,\boldsymbol{x}_{iL}\right]$), obtained from the data processing and embedding step, of
each function (e.g., $\bx_{i}$ can be from the source domain or the
target domain) is then used as the input to deep
learning models (e.g., the baselines and our proposed method).

Note that as the baselines, to our proposed method, for handling
the sequential properties of the data and to learn the automatic features
of the source code functions, we also use a bidirectional recurrent
neural network (bidirectional RNN) for both the source and target domains.

\vspace{1mm}
\subsection{Model configuration}
For the baselines including VulDeePecker \cite{VulDeePecker2018},
and DDAN \cite{Ganin2015}, MMD \cite{long2015}, D2GAN \cite{duald-tunguyen2017},
DIRT-T \cite{shu2018a}, HoMM \cite{Hommchen2020}, LAMDA \cite{LAMDAle21a},
SCDAN \cite{vannguyen2019dan} using the architecture CDAN proposed
in \cite{vannguyen2019dan}, and Dual-GD-DDAN and Dual-GD-SDDAN \cite{van-nguyen-dual-dan-2020},
and our proposed DAM2P method, we use one bidirectional recurrent
neural network with LSTM \cite{HochSchm97} cells where the size of
hidden states is in $\{128,256\}$ for the generator $G$ while
to the source classifier $C$ used in the baselines and the domain
discriminator $D$, we use deep feed-forward neural networks consisting
of two hidden layers where the size of each hidden layer is equal to $300$.
We embed the statement information in the $150$
dimensional embedding space. 

To our proposed method, the trade-off hyper-parameters
$\lambda$ and $\alpha$ are in $\{10^{-3},10^{-2},10^{-1}\}$ and $\{10^{-2},10^{-1},10^{0}\}$, respectively,
while the hidden size $h$ is in $\{128,256\}$. The dimension
of random feature space $2K$ is set equal to $1024$. The length $L$
of each function is padded or cut to 100 or less than 100 code statements (i.e., We
base on the quantile values of the functions\textquoteright{} length
of each dataset to decide the length of each function). We observe that almost all important information relevant to the vulnerability lies in the 100 first code statements or even lies in some very first code statements.

We employed the Adam optimizer \cite{KingmaB14} with an initial learning
rate of $10^{-3}$ while the mini-batch size is set to $100$ to our
proposed method and baselines. We split the data of the source domain
into two random partitions containing 80\% for training and 20\% for
validation. We also split the data of the target domain into two random
partitions. The first partition contains 80\% for training the models
of MMD, D2GAN, DIRT-T, HoMM, LAMDA, DDAN, SCDAN, Dual-GD-DDAN, Dual-GD-SDDAN,
and DAM2P without using any label information while the second partition
contains 20\% for testing the models. We additionally applied gradient
clipping regularization to prevent the over-fitting problem in the
training process of each model. For each method, we ran the corresponding
model $5$ times and reported the averaged measures. We implemented all mentioned methods
in Python using Tensorflow \cite{abadi2016tensorflow}, an open-source
software library for Machine Intelligence developed by the Google
Brain Team, on an Intel E5-2680, having 12 CPU Cores at 2.5 GHz with 128GB RAM, integrated NVIDIA Tesla K80.

\begin{figure}[h]
\begin{centering}
\vspace{-0mm}
\includegraphics[width=0.8\columnwidth]{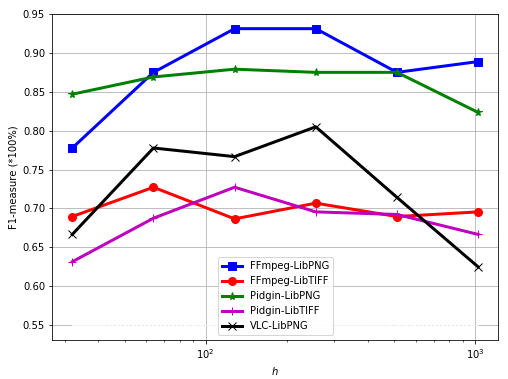}\vspace{-2mm}
\par\end{centering}
\caption{The correlation between $h$ and F1-measure of our proposed DAM2P
method. \label{fig:visual_peg_png_labels_ablation_study_h}}\vspace{-2mm}
\end{figure}

\vspace{1mm}
\subsection{Additional experiments}
\vspace{1mm}
\subsubsection{Hyper-parameter Sensitivity}

In this section, we investigate the correlation between important
hyper-parameters (including the $\lambda$, $\alpha$, and $h$ (the
size of hidden states in the bidirectional neural network)) and the
F1-measure of our proposed DAM2P method. As mentioned in the experiments
section, the trade-off hyper-parameters $\lambda$ and $\alpha$ are in
$\{10^{-3},10^{-2},10^{-1}\}$ and $\{10^{-2},10^{-1},10^{0}\}$, respectively, while the hidden size $h$ is in $\{128,256\}$.
It is worth noting that we use the commonly used values for the trade-off
hyper-parameters ($\lambda$ and $\alpha$) representing for the weights
of different terms mentioned in Eq. (6) and the hidden size $h$.
In order to study the impact of the hyper-parameters on the performance
of the DAM2P method, we use a wider range of values for $\lambda$,
$\alpha$, and $h$. In this ablation study, the trade-off parameters
$\lambda$ and $\alpha$ are in $\{10^{-4},10^{-3},10^{-2},10^{-1},10^{0},10^{1}\}$
while the hidden size $h$ is in $\{32,64,128,256,512,1024\}$.

\begin{figure}[h]
\begin{centering}
\vspace{-0mm}
\includegraphics[width=0.8\columnwidth]{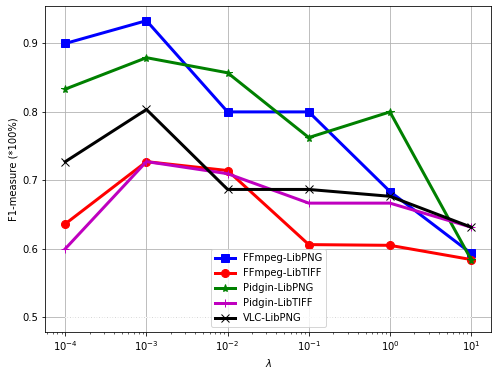}\vspace{-2mm}
\par\end{centering}
\caption{The correlation between $\lambda$ and F1-measure of our proposed
DAM2P method. \label{fig:visual_peg_png_labels_ablation_study_alpha}}\vspace{-2mm}
\end{figure}

\begin{figure}[h]
\begin{centering}
\vspace{-0mm}
\includegraphics[width=0.8\columnwidth]{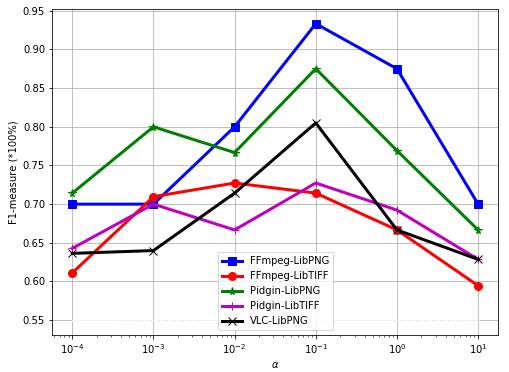}\vspace{-2mm}
\par\end{centering}
\caption{The correlation between $\alpha$ and F1-measure of our proposed DAM2P
method. \label{fig:visual_peg_png_labels_ablation_study_beta}}\vspace{-2mm}
\end{figure}

We investigate the impact of $\lambda,\alpha$, and $h$ hyper-parameters
on the performance of the DAM2P method on five pairs of the source
and target domains including FFmpeg to LibPNG, FFmpeg to LibTIFF,
Pidgin to LibPNG, Pidgin to LibTIFF, and VLC to LibPNG. As shown in
Figures (\ref{fig:visual_peg_png_labels_ablation_study_h}, \ref{fig:visual_peg_png_labels_ablation_study_alpha},
and \ref{fig:visual_peg_png_labels_ablation_study_beta}), we observe
that the appropriate values to the hyper-parameters used in the DAM2P
model in order to obtain the best model's performance should be in
from $10^{-4}$ to $10^{-2}$, from $10^{-3}$ to $10^{-1}$, and
from $64$ to $256$ for $\lambda,\alpha$, and $h$ respectively.
In particular, for the hidden size $h$, if we use too small values
(e.g., $\leq32$) or too high values (e.g., $\geq1024$), the model
might encounter the underfitting or overfitting problems respectively.
The model's performance on $\lambda$ (i.e., representing the weight
of the information from the target domain contributing to the cross-domain
kernel classifier during the training process) shows that we should
not set the value of $\lambda$ equal or higher than $1.0$ (i.e.,
used for the weight of the information from the source domain), and
the value of $\lambda$ should be higher than $10^{-4}$ to make sure
that we use enough information of the target domain in the training
process to improve the cross-domain kernel classifier.

\end{document}